\documentclass{jfm}

\usepackage{soul}
\usepackage{graphicx}
\usepackage{newtxtext}
\usepackage{newtxmath}
\usepackage{natbib}
\usepackage[]{amsmath}

\usepackage{booktabs}
\usepackage{subfig}
\usepackage{floatrow}
\usepackage{upgreek}
\usepackage{overpic}
\usepackage[]{multirow}
\usepackage[]{url}
\usepackage[]{nicefrac}
\usepackage{bm}
\usepackage{wasysym}
\usepackage{enumerate}
\usepackage{diagbox}
\usepackage{algorithm}
\usepackage{algpseudocode}
\usepackage{hyperref}
\hypersetup{
    colorlinks = true,
    urlcolor   = blue,
    citecolor  = black,
}

\newcommand{\RomanNumeralCaps}[1]
\linenumbers

\newlength\myboxwidth
\begin{document}

\title{
Jet mixing enhancement with Bayesian optimization, deep learning, and persistent data topology
}
\author{Yiqing Li\aff{1}, 
         Bernd R. Noack\aff{1,2}\corresp{\email{bernd.noack@hit.edu.cn}},
        Tianyu Wang\aff{1}, 
        Guy Y. Cornejo Maceda\aff{1}\corresp{\email{yoslan@hit.edu.cn}},\\
        Ethan Pickering\aff{3}, 
        Tamir Shaqarin\aff{4},
        and Artur Tyliszczak\aff{5}
        }
\affiliation{
\aff{1} Chair of Artificial Intelligence and Aerodynamics, 
School of Mechanical Engineering and Automation, Harbin Institute of Technology, 518055 Shenzhen, P.~R.~China
\aff{2} 
Guangdong Provincial Key Laboratory of Intelligent Morphing Mechanisms and Adaptive Robotics,
Harbin Institute of Technology, 518055 Shenzhen, P.~R.~China
\aff{3} Independent Scholar
\aff{4} Department of Mechanical Engineering, Tafila Technical University, 
66110 Tafila, Jordan
\aff{5} Faculty of Mechanical Engineering and Computer Science, Czestochowa University of Technology, 42-201 Czestochowa, Poland
}


\maketitle


\begin{abstract}
    
    We optimize the jet mixing using large eddy simulations (LES) at a Reynolds number of $3000$.
    Key methodological enablers consist of Bayesian optimization, a surrogate model enhanced by deep learning, and persistent data topology for physical interpretation.
    The mixing performance is characterized by an equivalent jet radius ($R_{\rm eq}$) derived from the streamwise velocity in a plane located $8$ diameters downstream.
    The optimization is performed in a 22-dimensional actuation space that comprises most known excitations.
    The plant benefits from a 22-dimensional actuation space that comprises most known excitations.
    This search space parameterizes distributed actuation imposed on the bulk flow and at the periphery of the nozzle in the streamwise and radial directions.
    The momentum flux measures the energy input of the actuation.
    The optimization quadruples the jet radius $R_{\rm eq}$
    with a $7$-armed blooming jet after around $570$ evaluations.
    The control input requires  $2\%$ momentum flux of the main flow, which is one order of magnitude lower than an ad hoc dual-mode excitation.
    Intriguingly, a pronounced suboptimum in the search space 
    is associated with a double-helix jet, a new flow pattern.
    This jet pattern results in a mixing improvement comparable to the blooming jet.
    A state-of-the-art Bayesian optimization converges
    towards this double helix solution.
    The learning is accelerated and converges to another better optimum by including surrogate model trained along the optimization. Persistent data topology extracts the global and many local minima in the actuation space.
    These minima can be identified with flow patterns beneficial to the mixing.
    \end{abstract}
\section{Introduction}
\label{ToC:Introduction}
    Jet flows are ubiquitous in nature and technology
    and belong to a handful of configurations described in any fluid mechanics textbook. Jet mixing plays a pivotal role in many engineering applications, e.g.\ fuel injection in engines, combustor cooling, chemical mixing, printing, and noise generation \citep{Jordan2013arfm}, just to name a few.
    Hence, jet mixing optimization plays an important part in academic research and engineering applications.

    Laminar jets are affected by the Kelvin-Helmholtz instability of the initial shear layer \citep{ball2012pas}.
    The jet shear layer rolls up into pronounced vortex rings.
    Excitation at the nozzle exit provides authority over the vortex formation, e.g.\ allows to speed up the vortex formation,  to promote or mitigate vortex pairing, 
    and to influence the far-field coherent structures.
    Vortex pairing in streamwise direction promotes larger mixing regions 
    observed as the orderly ‘vortical puffs’ with axisymmetric excitation \citep{crow1971jfm}.
    More importantly,
    a significant increase in the spreading angle can be obtained by vortex splitting evolving along several branches \citep{lee1985reporttf}. 

    The actuation may promote axisymmetric, helical, dual-mode, flapping, and bifurcating dynamics. 
    In particular,  acoustic excitation of  bulk affects the jet spreading
    via controlled vortex pairing \citep{crow1971jfm,hussain1980jfm}.
    Jet mixing is more effectively augmented with helical forcing 
    \citep{mankbadi1981ptrs, corke1993jfm}.
    Bifurcating, 
    trifurcating, 
    and blooming jets 
    appear with a spreading angle up to $80^\circ$ 
    when axisymmetric and helical modes are combined (dual-mode) with different frequency ratios 
    \citep{lee1985reporttf}.
    The flapping mode is composed of counter-rotating helical modes, and the combination of axisymmetric and flapping modes is referred to as bifurcating mode.
    Both the flapping and the bifurcating mode can produce bifurcating jets with an impressive jet spreading \citep{parekh1989phd, danaila2000pf, da2002pf}.

    The world of multiple-mode actuation for mixing optimization
    holds considerable promise and is still to be explored.
    The radial excitation with three flapping modes, including $9$ parameters, is optimized by Evolution Strategies \citep{Koumoutsakos2001aiaaj}. Only one dominant flapping mode remains after $400$ Direct Numerical Simulations (DNS) at $Re=500$. 
    In the sequel, 
    the bifurcating mode using axial forcing is optimized 
    for $Re$  up to $1500$ using
    the amplitudes and two Strouhal numbers as control parameters \citep{hilgers2001fdr}.
    In an adjoint-based optimization study  at $Re=2000$, 
    radial forcing is found to be more effective than axial actuation in dual-mode forcing 
    \citep{shaabani2020jfm}.
    In an experiment at $Re=8000$,
    jet mixing is manipulated with periodic operation 
    of six radial minijets.
    In 200 evaluations, 
    Bayesian optimization minimizes a streamwise centerline velocity 
    tuning 12 parameters, the frequency, amplitudes, and phase differences \citep{blanchard2021ams}.
    The optimal mixing is facilitated by combining flapping and helical forcing,
    like machine learning control for the same configuration \citep{zhou2020jfm}.
    Moreover, the control performance also benefits from the deployment of more actuators and richer actuation space. 
    For example, an intelligent nozzle with eighteen electromagnetic flap actuators \citep{suzuki1999symposium}, and 8-channel localized arc filament plasma actuators \citep{utkin2006jfd} have been developed for jet control.

    In flow control, machine learning techniques have recently gained attention due to their successful applications \citep{Duriez2017book, Brunton2020arfm}. 
    Examples are genetic programming and variants \citep{Cornejo2021jfm}, reinforcement learning \citep{rabault2019jfm, nair2023jfmr,Sonoda2023jfm,Vignon2023pof_marl,Vignon2023pof,Xu2023jfm,Guastoni2023epje}, and Bayesian optimization \citep{blanchard2021ams}.
    These methods encode the input-output relations in various forms without requiring prior knowledge.
    Function regression solvers like genetic programming and deep reinforcement learning can provide a large model capacity for exploration.
    However, deriving the optimal solution in finite time can not be guaranteed.
    Alternatively, a predefined control law can be tuned to near-optimal by parameter optimizers like Bayesian optimization (BO), genetic algorithm (GA) and particle swarm optimization (PSO), to name a few.
    \citet{pino2023jfm} compares genetic programming, deep reinforcement learning, and BO in 
    increasingly complex control problems. 
    The authors highlight BO's potential to balance both sample efficiency and the performance of the final solution.
    With the recent advances in the design of the acquisition function \citep{blanchard2021jcp} and the surrogate models \citep{ pickering2022ncs},
    BO is moving forward in conquering high-dimensional search spaces.
    This work leverages these advancements to optimize and understand high-dimensional jet forcing modes.

    The present study builds on a jet mixing plant employing Large Eddy Simulations (LES)
    and a rich streamwise and radial actuation space at the nozzle exit.
    This plant can reproduce virtually all previously considered actuated jet dynamics
    as elements of a high-dimensional search space.
    High-dimensional optimization constitutes a challenge
    that is tackled by a Bayesian optimizer enhanced by deep learning.

    The paper is organized as follows.
    The configuration, actuation and metrics are defined in \S~\ref{ToC:setup}.
    The optimizer and numerical solver are presented in \S~\ref{ToC:method}. We discuss the learning process and the optimized solutions in \S~\ref{ToC:res}.
    Finally, \S~\ref{ToC:conclusion} concludes the findings with outlook.
\section{Setup and problem definition}
\label{ToC:setup}
    \subsection{Configuration and actuation}
    \label{ToC:setup:b}
    \begin{figure}[htb]
        \centering
        \includegraphics[width=0.8\textwidth]{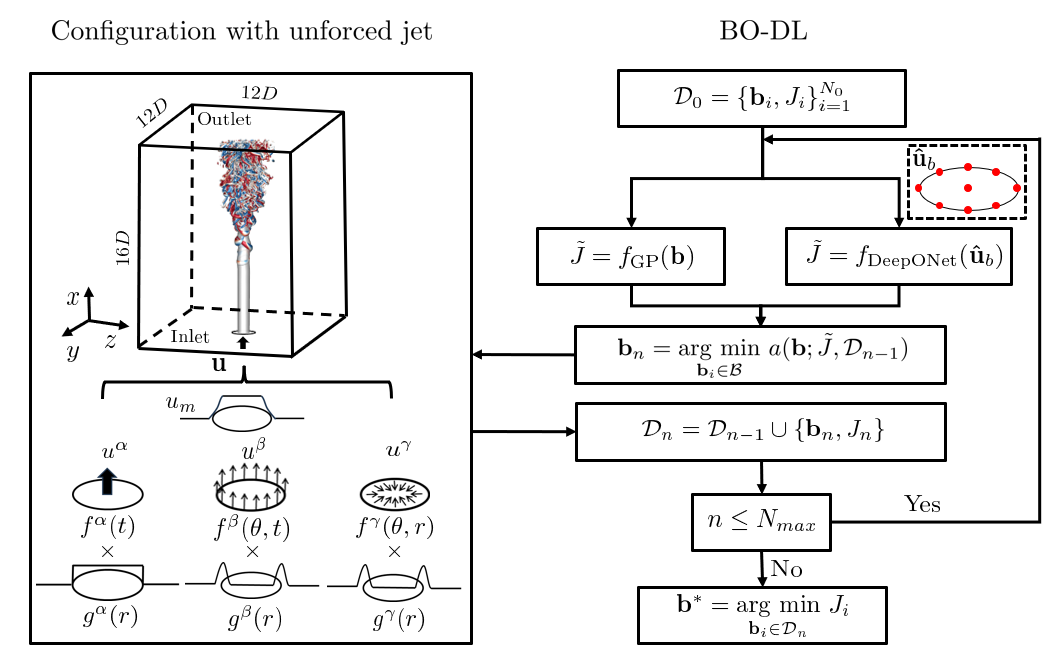}
        \caption{Problem setup including the jet configuration with the designed actuation (left) and the deep learning enhanced Bayesian optimization (right).
        $\mathbf{b}$ is the actuation parameter.
        $\hat{\mathbf{u}}_b$ is the actuation command at the 9 red points indicated in the dashed box.
        }
        \label{fig:framework}
    \end{figure}
    The configuration is a jet flow exiting a circular nozzle of diameter $D$ in figure~\ref{fig:framework}.
    The flow is described in a Cartesian coordinate system $(x,y,z)$
    where $x$ represents the streamwise direction
    and the origin coincides with the center of the nozzle.
    The computational domain starts from the exit and covers a rectangular region with size $12D\times16D\times12D$. 
    The actuation $\bm{u}_b(r, \theta, t)$ is imposed with the mean streamwise velocity $u_m(r)$ as the inlet velocity profile $u(r, t)$, $$\bm{u}(r,t) = u_m(r) \bm{e}_x + \bm{u}_b(r, \theta, t),$$ 
    where $r$ measures the radial distance from the centerline, 
    and $\theta$ is the azimuthal angle.
    The mean streamwise component has a hyperbolic-tangent profile,
    $$ u_m(r) = \frac{U_{\rm j}+u_c}{2}-\frac{U_{\rm j}-u_c}{2}\tanh\left(\frac{1}{4}\frac{R}{\delta_2}\left(\frac{r}{R}-\frac{R}{r}\right)\right),$$
    where $U_{\rm j}$ is the jet centerline velocity, $u_{c}=0.03 U_{\rm j}$ is the co-flow velocity to mimic a natural suction process, and $\delta_2=R/20$ is the momentum boundary layer thickness of the initial shear layer.
    At the side boundaries, we impose the vertical velocity equals $u_c$, and the remaining velocity components equal zero. The pressure at the side boundaries is computed from the Neumann condition ${\mathbf{n}} \cdot \nabla p=0$ with ${\mathbf{n}}$ as the vector normal to the boundary. At the outlet plane, the velocity is computed from a convective boundary condition $\partial \bm{u}/\partial t + \tilde{V}_C \partial \bm{u}/\partial n=0$, where $\tilde{V}_C$ is the instantaneous convection velocity $V_C$ limited to positive values: $\tilde{V}_C=\max(V_C,0)$. $V_C$ is the velocity averaged over the outlet plane.
    The pressure at the outflow equals zero. Such defined outflow boundary condition ensures stable simulations and has negligible impact on the turbulent flow structures leaving the computational domain \citep{Tyliszczak2014ftc,Tyliszczak2018ijhff}.

    As introduced in \S~\ref{ToC:Introduction}, 
    the jet control techniques for mixing enhancement are usually designed according to the instability mode, described by their azimuthal wavenumber at order $0$ (axisymmetric mode) or $1$ (helical mode).
    The perturbation is either axial or radial.
    We combine both axial and radial perturbation and define the actuation $\bm{u}_b$ with a general expression of $\theta$ and $t$, without assumption on the forcing mode. 
    Therefore, the term $\bm{u}_b$ includes an axisymmetric streamwise bulk forcing $u^{\alpha}(r,t)$, and a peripheral forcing with the streamwise component $u^{\beta}(r,\theta,t)$, and the radial $u^{\gamma}(r,\theta,t)$:
    \begin{equation}
        \label{eq:u}
        \bm{u}_b =  \left(u^\alpha
                                + u^\beta \right) \bm{e}_x
                                + u^\gamma \bm{e}_r.
    \end{equation}

    The forcing components are the product of a perturbation $f^m(\theta,t)$ and a radial profile $g^m(r)$: $u^m(r,\theta,t) = f^m(\theta,t)g^m(r)$, $m=\alpha, \beta, \gamma$ with $g^{\alpha}(r)=1$ for $r \leq R$ and $0$ for $r > R$, and $g^{\beta}(r) = g^{\gamma}(r) = \exp(-1000(R-r)^{2.5})$.
    The profiles of the three forcing components are depicted in figure~\ref{fig:framework}.
    The perturbation terms $f^m(\theta,t)$ are defined as the sum and product of space- and time-harmonic functions:
    \begin{eqnarray}\label{eq:falpha}
        f^\alpha (t) &=& \sum_{i=-L}^L \alpha_i  \> \Theta_i( \omega_\alpha t) \\
        f^\beta(\theta, t)  &=& \sum _{i,j=-L, \ldots, L} \beta_{ij} \> \Theta_i(\theta) \> \Theta_j(\omega_\beta t) \label{eq:fbeta} \\
        f^\gamma(\theta, t)  &=& \sum _{i,j=-L, \ldots, L} \gamma_{ij} \> \Theta_i(\theta) \> \Theta_j(\omega_\gamma t), \label{eq:fgamma}
    \end{eqnarray}
    where $\alpha_i$, $\beta_{ij}$, $\gamma_{ij}$
    and $\omega_{\alpha}$, $\omega_{\beta}$, $\omega_{\gamma}$ 
    are the actuation amplitudes and angular frequencies, respectively.
    $\Theta_i(\phi)$ is the harmonic function basis defined as: $\Theta_i(\phi)=\sin (i \phi)$ for $i > 0$,  $\Theta_i(\phi)=1$ for $i = 0$, and $\Theta_i(\phi)=\cos (i \phi)$ for $i < 0$.
    The forcing ansatz can approximate 
    any periodic function of $\theta$ and $t$ as the expansion order increases.
    In this study, focus is placed on the 
    first order expansion ($L=1$) of (\ref{eq:falpha}-\ref{eq:fgamma}). 
    Thus, the control law is parameterized by 
    a $22$-dimensional vector $\mathbf{b}$,
    \begin{equation}
        \label{equ:b}
        \mathbf{b} = \left[ St_\alpha, St_\beta, St_\gamma, \alpha_1, 
        \left \{ \beta_{ij} \right\}_{i,j=-1,0,-1},
        \left \{ \gamma_{ij} \right\}_{i,j=-1,0,-1}\right]^{\intercal} \in \mathcal{B},
    \end{equation} 
    where the Strouhal numbers $St_m = \omega_m D/2 \pi U_{\rm j}, m=\alpha,\beta,\gamma$.
    Note that $\alpha_0$ is set to 0
    as a constant bulk flow can be incorporated into the steady profile.
    In addition, $\alpha_{-1} = 0$ can be assumed by a translation in time.
    The range of $St_m$ is set as $[0.1,1]$ to include the Strouhal number of the preferred mode at $St_p=0.3-0.64$ \citep{crow1971jfm,GutmarkHo_PoF_1983,Sadeghi_Pollard_PoF_2012}, and the range of axisymmetric mode $St_\alpha \in [0.15,0.8]$ where bifurcating and blooming jets are observed \citep{lee1985reporttf, parekh1989phd,Tyliszczak2018ijhff}.
    The actuation amplitudes are limited to $-0.1 \le \alpha_1, \beta_{ij}, \gamma_{ij}    \le  0.1$, lower than $0.15$ used by \cite{danaila2000pf,gohil2015jfm,Tyliszczak2018ijhff}, and $0.5$ by \cite{Koumoutsakos2001aiaaj}.

    This high-dimensional search space allows the actuation to emulate various forcing modes, such as axisymmetric, helical, flapping, bifurcating, dual-mode, and harmonic waves discussed in \S~ \ref{ToC:Introduction}. In short,
    the forcing can be axisymmetric, statistically axisymmetric, and non-axisymmetric.
    \subsection{Mixing and actuation performance}
    \label{ToC:setup:radius}
    The mixing process in turbulent jets is typically characterized by quantities such as the decay of centerline velocity and its fluctuations, or the entrainment \citep{nathan2006prog}.
    However,
    these quantities only measure the local statistics and cannot reflect the mixing process of 
    non-axisymmetric jets, such as bifurcating jets and asymmetric jets.
    For example, the velocity in these jets may locally drop to zero due to a jet-splitting phenomenon, rather than as a result of enhanced mixing.
    The entrainment is rather a measure of the amount of surrounding fluid entrained into the jet vicinity,
    without guaranteeing that it mixes with the
    jet.
    In asymmetric jets, the amount of fluid flowing towards the jet may characterize significant radial non-uniformity not revealed by the entrainment.
    Considering the non-axisymmetric forcing in search spaces, 
    we define a new metric, the equivalent mixing radius $R_{\rm eq}$, to estimate the spatial uniformity of the streamwise velocity.
    $R_{\rm eq}$ is defined as the normalized streamwise velocity variance computed at a given $y-z$ cross-section:
    \begin{equation}\label{eq:radius_eq}
    R_{\rm eq}=\sqrt{2} \sigma, ~
    \sigma^2  = \frac{\int \! \! \! \int \! 
    \varrho (y,z)  \> \left[(y-y_c)^2 + (z-z_c)^2  \right] dydz}{\int \! \! \!\int\varrho (y,z)  dydz  },
    \end{equation}
    with $\varrho(y,z)=(\langle{u(x=X_0,y,z)}\rangle_t-u_c)/(U_{\rm j}-u_c)$,
    $X_0=8D$, and $(y_c, z_c)$ the jet center, as an analog to the center of mass:
    $y_c = \int \! \! \!\int y \varrho (y,z)  dydz / \int \! \! \!\int\varrho (y,z)  dydz$
    and
    $z_c = \int \! \! \!\int z \varrho (y,z)  dydz / \int \! \! \!\int\varrho (y,z)  dydz $.
    The $\sqrt{2}$ coefficient is chosen so that the equivalent mixing radius of a top flat jet flow of radius $R$ is $R$.

    The amplitude and mass flow rate have been adopted to evaluate the control input.
    Inspired by \cite{parekh1989phd},
    we define the momentum flux $P$ of the actuation to estimate the energy input from a practical perspective.
    The momentum flux is time-averaged and normalized by the jet axial momentum flux at the inlet:
    \begin{equation}
        P=\frac{\langle{\int \! \! \!\int u^2_b dA}\rangle_t}{\pi \> R^2 \> U_{\rm j}^2}.
    \end{equation}
\section{Methodology}
\label{ToC:method}
    \subsection{Deep learning-enhanced Bayesian optimization}
    \label{ToC:method:bo}
    The optimization problem to maximize the mixing as a response to the actuation input parameterized by $\mathbf{b}$ is formulated as
    \begin{equation}
        \label{eq:Min_problem}
        \mathbf{b}^* = \underset{\mathbf{b}\in \mathcal{B}}{\arg \min} \; J(\mathbf{b}),
    \end{equation}
    where $\mathcal{B} = [0.1,1]^3\times [-0.1,0.1]^{19}  \subset \mathbb{R}^{22}$. Cost function $J$ is defined as the inverse of the equivalent mixing radius and normalized by the unforced case, $J=R_{\rm eq,0}/R_{\rm eq}$.
    Better mixing with large $R_{\rm eq}$ is related to the decrease of $J$. The assumed optimization goal leads to the spatial uniformity of $\langle{u(x=X_0,y,z)}\rangle_t$.

    To optimize this $22$-dimensional search space,
    we employ techniques inspired by Bayesian optimization (BO) \citep{williams2006mitbook}. BO has shown to be advantageous in optimizing expensive black-box functions by systematically reducing uncertainty in the black-box mapping and incorporating prior assumptions of the cost function \citep{shahriari2015ieee}.
    Through a sequential approach, BO identifies the next actuation to evaluate, or ``data point'' to acquire, for the purpose of finding the optimum. This is generally achieved via a surrogate model trained on all the queried data and an acquisition function \citep{williams2006mitbook}.
    A sketch of the method used in this study is shown in figure~\ref{fig:framework}.
    The algorithm is initialized with the evaluation of a set $\mathcal{D}_0$ of $N_0$ actuation vectors in $\mathcal{B}$ generated by Latin hypercube sampling. $N_0$ is equal to $N_D+1$ with $N_D$ the dimension of the search space $\mathcal{B}$.
    We recall that for this study, $N_D=22$.
    $\mathcal{D}_0$ includes all the evaluated parameter vectors and their cost $\{\mathbf{b}_i, J_i \}_{i=1}^{N_0}$.
    A surrogate model $\tilde J$ is trained on the available data 
    to approximate the latent objective function $J$.
    After the initialization, 
    the algorithm explores the search space $\mathcal{B}$ one new query at a time. 
    At each iteration, BO determines the optimal actuation to implement next by minimizing an acquisition function $a(\mathbf{b})$.
    The acquisition function leverages the surrogate model $\tilde J$ and available data $\mathcal{D}_{n-1}$ 
    to guide the data selection in the search space. 
    After each query, the data set is enriched by the new data point $\{\mathbf{b}_n, J_n\}$ into $\mathcal{D}_{n}$ to further refine the surrogate model.
    When the query budget is met, 
    the algorithm ends with the best design vector $\mathbf{b}^*$ recorded during the optimization.

    The two key elements in BO are the choice of the surrogate model and the sequential strategy \citep{blanchard2021jcp}.
    We focus on the former for a better estimation of the high-dimensional flow control system.
    Gaussian processes (GP) serve as a successful surrogate model in moderate dimensions and can provide closed-form solutions with the posterior distribution.
    However, the computation of the posterior costs $\mathcal{O}(n^3)$ where $n$ is the number of observations due to the inverse of the covariance matrix.
    The number of evaluations required to effectively cover the domain grows exponentially with the dimensionality.
    This makes GP difficult to scale to large training sets for high-dimensional problems.
    Recently, the deep operator network (DeepONet) has shown small generalization error for systems where functions are acted upon by an operator \citep{lu2021nmi}.
    Different from GP that parameterizes the input, 
    DeepONet can map input functions, which are then transformed by an operator, to an output function or scalar with improved accuracy.
    This means that DeepONet does not fall victim to the scaling difficulties of GP when training. Therefore, DeepONet is capable of learning from infinite-dimensional functions.
    Empirically, DeepONet's utility as an operator surrogate model for Bayesian-inspired experimental design has been shown to significantly outperform GP in several infinite-dimensional systems that exhibit extreme events in \cite{pickering2022ncs}, ranging from stochastic pandemic spikes, to catastrophic structural failure, to rogue wave identification.
    Through a study of the Bayesian optimizer based on GP (BO) and DeepONet (BO-DeepONet) for the defined 22-dimensional problem (see \S \ref{ToC:res:bos}),
    we propose a new algorithm, deep learning-enhanced Bayesian optimization (BO-DL).
    By incorporating both GP and DeepONet as the surrogate model,
    BO-DL presents a better explorative capability and faster convergence. 
    We alternate between DeepONet and GP every $10$ iterations to query the next sample.
    The $10$ is chosen empirically to balance the characteristics of the two models and combine their advantages.
    If the interval is too long, such as $100$, the models will be more independent than interacting with each other.
    On the other hand, if the interval is too short, such as $1$, the exploitation may be interrupted by uncertainty due to the exchange of models.
    In GP implementations, the parameter space of a stochastic process is used for both regression and searching. Instead, DeepONet performs regression in the functional space, leveraging the typically disregarded basis functions associated with the parameterization.
    Here, the function $\hat{\mathbf{u}}_b$ is designed as the actuation command at $9$ points located at the jet exit, including the centerline and $8$ equidistant points on the periphery.

    The acquisition function employed is the likelihood-weighted lower confidence bound proposed by \cite{blanchard2021jcp}, with superiority in finding rare extreme behaviors,
    \begin{equation}
        \label{equ:acquisition}
        a(\mathbf{b}) = \mu(\mathbf{b}) - \kappa \sigma(\mathbf{b}) w(\mathbf{b}), ~ w(\mathbf{b}) = \frac{p_\mathbf{b}(\mathbf{b})}{p_\mu(\mu(\mathbf{b}))}.
    \end{equation} 
    Here, $\kappa$ balances exploration (large $\kappa$) and exploitation (small $\kappa$),
    and is chosen as $1$.
    The mean model $\mu(\mathbf{b})$ and standard variance model $\sigma(\mathbf{b})$ 
    are estimated by the mean and variance over a $2$-ensemble of trained DeepONet. 
    For the case of GP, these can be calculated in closed form using standard expressions from GP regression \citep{williams2006mitbook}. 
    The likelihood ratio $w(\mathbf{b})$ measures relevance by weighting the uncertainty of the point (the input density $p_\mathbf{b}$) 
    against its expected impact on the cost function (the output density $p_\mu$).

    \subsection{Governing equations and numerical solver}
    \label{ToC:method:les}
    We consider an incompressible flow described by the Navier-Stokes equations in the framework of LES:
    \begin{equation}\label{eqn:CONT_LES}
    \frac{\partial \overline{ {u}}_j}{\partial x_j}=0,
    \end{equation}
    \begin{equation}\label{eqn:NS_LES}
    \frac{\partial \overline{u}_i}{\partial t}+\frac{\partial  {\overline{u}_i \overline{u}_j}}{\partial x_j}=-\frac{1}{{\rho}}\frac{\partial \bar{p} }{\partial x_i}+\frac{\partial \tau_{ij}}{\partial x_j} + \frac{\partial \tau_{ij}^{f}}{\partial x_j},
    \end{equation}
    where ${u}_i$  represents the velocity components, $p$ denotes pressure and $\rho$ is density. The overbar denotes spatially filtered variables, $\bar{f}(\mathbf{x},t)=\int_{\Omega}f(\mathbf{x}',t)\mathcal{G}(\mathbf{x}-\mathbf{x}',\Delta) d\mathbf{x}'$ and $\mathcal{G}$ is the filter function that fulfils the condition $\int_{\Omega}\mathcal{G}(\mathbf{x},\Delta) d\mathbf{x}=1$. 
    A local filter width equals the cube root of the computational cell volume, $\Delta=(\Delta x\Delta y\Delta z)^{1/3}$. 
    The stress tensor includes the large-scale term $\tau_{ij}$ and the sub-grid term $\tau_{ij}^{f}$ defined as
    \begin{equation}
    \tau_{ij}=2\nu S_{ij},\quad \tau_{ij}^{f}=\left({\overline{u}_i \overline{u}_j} - \overline{{u}_i{u}_j}\right),
    \end{equation}
    where $\nu$ is the kinematic viscosity and $S_{ij}=\frac{1}{2}\left(\frac{\partial \bar{u}_i}{\partial x_j} + \frac{\partial \bar{u}_j}{\partial x_i}\right)$ is the rate of strain tensor of the resolved velocity field.
    In this work, the sub-filter tensor is modeled by an eddy-viscosity model with $\tau_{ij}^{f} = 2\nu_tS_{ij} + \tau^f_{kk}\delta_{ij}/3$. The diagonal terms $\tau^f_{kk}$ are added to the pressure, resulting in the so-called modified pressure $\bar{P}=\bar{p}-\rho\tau^f_{kk}\delta_{ij}/3$.
    The Vreman subgrid-scale model 
    is used for the low computational cost and very good accuracy in simulating jet flows \citep{Wawrzak_Boguslawski_Tyliszczak_FTaC_2015,boguslawski2019jfm}.

    The simulations are conducted with an in-house high-order LES solver SAILOR. The solution algorithm is based on the projection method for the pressure-velocity coupling for half-staggered meshes where the pressure nodes are shifted half a cell size from the velocity nodes \citep{tyliszczak2014jcp,tyliszczak2015cf}. A predictor-corrector method (Adams-Bashforth/Adams-Moulton) is applied for the time integration. Derivative approximations and interpolation on staggered nodes are defined using 6$^{th}$ and 10$^{th}$ order compact difference formulas. 
    The SAILOR solver was used in the jet studies with similar dynamic scales as the present work, such as jets undergoing laminar/turbulent transition 
    \citep{boguslawski2019jfm} 
    and excited jets 
    \citep{Tyliszczak2014ftc,Tyliszczak2018ijhff}. 
    The applied high-order discretization schemes led to grid-independent results already at relatively coarse meshes. 
\section{Results}
\label{ToC:res}
    \subsection{LES validation with the bifurcating jet}
    The Reynolds number $\Rey = U_{\rm j} D/\nu$ is decided as $3000$ for this study.
    This allows for the use of a relatively coarse computational mesh to obtain reliable and fast simulations as the database. 
    Two meshes are employed in this study. A coarse mesh with $80\times160\times80$ nodes is used for the learning process
    and a refined mesh with $192\times336\times192$ nodes for the validation and flow analysis of selected cases. The mesh points are compacted in the axial direction towards the inlet using an exponential function and radially towards the jet axis by a tangent hyperbolic function. In the region $-1.2D<y,z<1.2D$, the mesh spacing is almost uniform and equal to $\Delta y=\Delta z=0.05D$ (46 nodes) on the coarse mesh and $\Delta y=\Delta z=0.02D$ (115 nodes) on the dense one. In the axial direction, the sizes of the cells in the direct inlet vicinity are $\Delta x=0.067D$ and $\Delta x= 0.032D$ for the coarse and dense mesh, respectively. The time step varies according to the Courant-Friedrichs-Lewy condition, with the CFL number equal to 0.5. 
    The jet impulsively injects into quiescent flow and becomes fully developed after $100 D/U_{\rm j}$ time units.
    The time-averaging procedure then starts and lasts for $500D/U_{\rm j}$ time units for the statistics to converge. 
    A single simulation on the coarse mesh takes $20$ CPU-hours. The whole optimization process with $1000$ converged simulations lasts around $21$ days, using 40 CPUs of an AMD EPYC 7742 (2.25GHz) processor. On the dense mesh, the single simulation run takes approximately $576$ CPU-hours.
    The parallel computation is carried out with the Open MPI interface.

\begin{figure}[ht]
        \centering
        \includegraphics[width=0.95\textwidth]{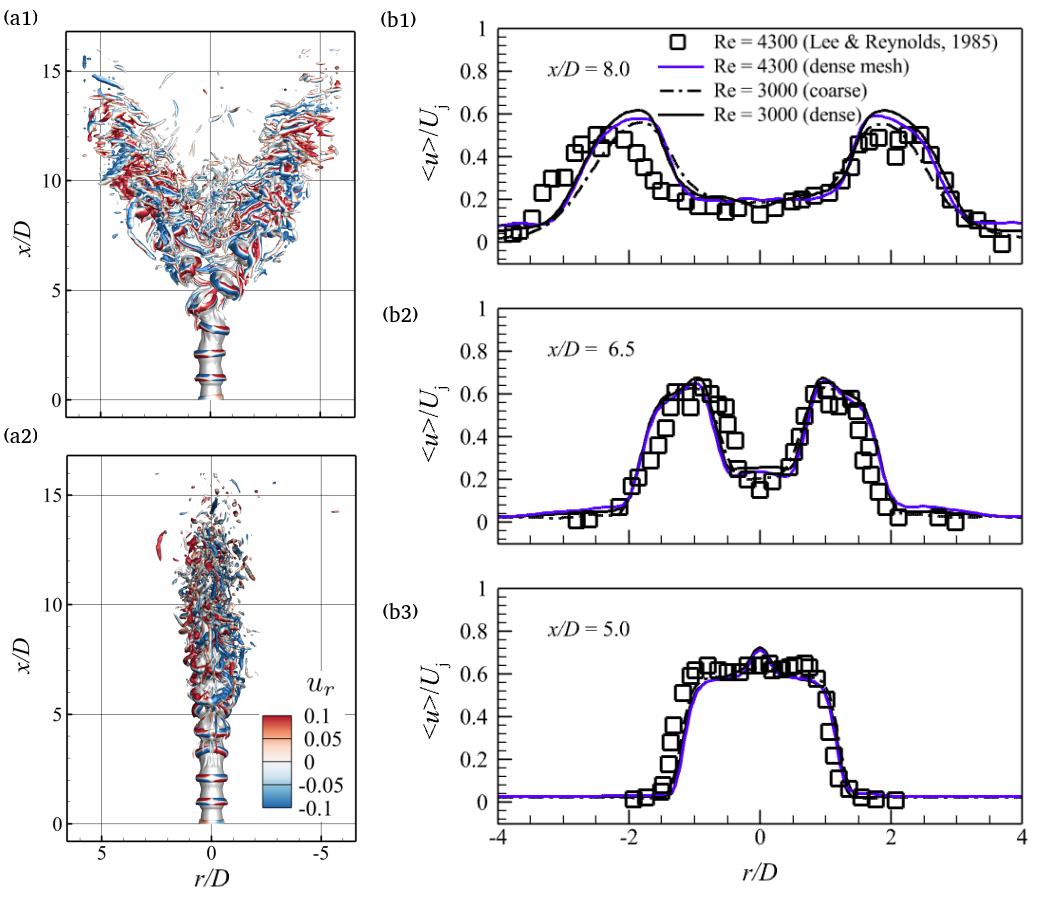}
        \caption{
        Validation of the LES solver on the bifurcating jet at $\Rey=4300$ \citep{lee1985reporttf}.
        Instantaneous isosurfaces of Q-parameter ($Q=0.5$) in the bifurcating (a1) and bisecting plane (a2), colored by the radial velocity $u_r$. Radial profiles of the time-averaged axial velocity in the bifurcating planes (b1-b3). The square symbols represent the experimental results of \citet{lee1985reporttf}, and the lines represent the simulation results obtained in this study.}
        \label{fig:bif-plane}
    \end{figure}

    As presented in \ref{ToC:method:les},
    the numerical code employed has been well validated against experimental and numerical data for a series of studies of jet dynamics and control.
    Here, the code with the assumed perturbation design~(\ref{eq:falpha}-\ref{eq:fgamma}) is verified to obtain a well-documented flow pattern of an excited jet,  
    the bifurcating jet at $\Rey=4300$ \citep{lee1985reporttf}. 
    The well-known jet is shown in figure~\ref{fig:bif-plane}, produced by
    the combined axisymmetric and flapping excitation at the frequencies $0.4<St_\alpha<0.6$ and $St_\beta=St_\gamma=St_\alpha/2$. 
    Based on the current control definition, the bifurcating jet is reproduced by
    \begin{eqnarray}\label{eq:falpha-bif}
        f^\alpha (t) = \alpha_{1}  \> \Theta_{1}( \omega_\alpha t)
    \end{eqnarray}
    that produces the axisymmetric excitation and
    \begin{eqnarray}\label{eq:falpha-bif2}
        f^\beta(\theta, t)  = \beta_{-1,1} \> \Theta_{-1}(\theta) \> \Theta_{1}(\omega_\beta t) \label{eq:fbeta-bif},\, \quad
        f^\gamma(\theta, t)  = \gamma_{-1,1} \> \Theta_{-1}(\theta) \> \Theta_{1}(\omega_\gamma t), \label{eq:fgamma-bif}
    \end{eqnarray}
    as the flapping mode, simulating the orbital motion of the nozzle tip in the experiment.
    We take $\alpha_{1}=0.17$, $St_\alpha=0.5$ and assume $St_\beta=St_\gamma=St_\alpha/2$, $\beta_{-1,1}^2+\gamma_{-1,1}^2=\alpha_{1}^2$ with $\beta_{-1,1}=\alpha_{1}\cos(20^\circ)$.
    This type of excitation is also used in the previous LES simulations of the bifurcating jet \citep{da2002pf,Tyliszczak2014ftc}. 

    Figure~\ref{fig:bif-plane}(b1-b3) shows the time-averaged axial velocity profiles along the radius in the bifurcating plane at the distance $x/D=5.0,\, 6.5,\, 8.0$ from the inlet. These results were obtained for $\Rey=4300$, as in \cite{lee1985reporttf}, and for $\Rey=3000$ assumed in the present study. 
    The effect of the Reynolds number on the velocity profiles is small.
    We attribute such behavior to a dominating role of the perturbation.
    The location and level of two peaks, which are associated with the split jet arms, are well predicted by the numerical solutions.
    The impact of the mesh density on the solution is also negligible, owing to the employed high-order numerical method.
    This also holds for the optimized jet, see figure \ref{fig:res:ucl} of \S~\ref{ToC:res:solutions}.
  
    \subsection{Bayesian Optimization with different surrogate models}
    \label{ToC:res:bos}
    \begin{figure}[htb]
        \centering
        \includegraphics[width=0.9\textwidth]{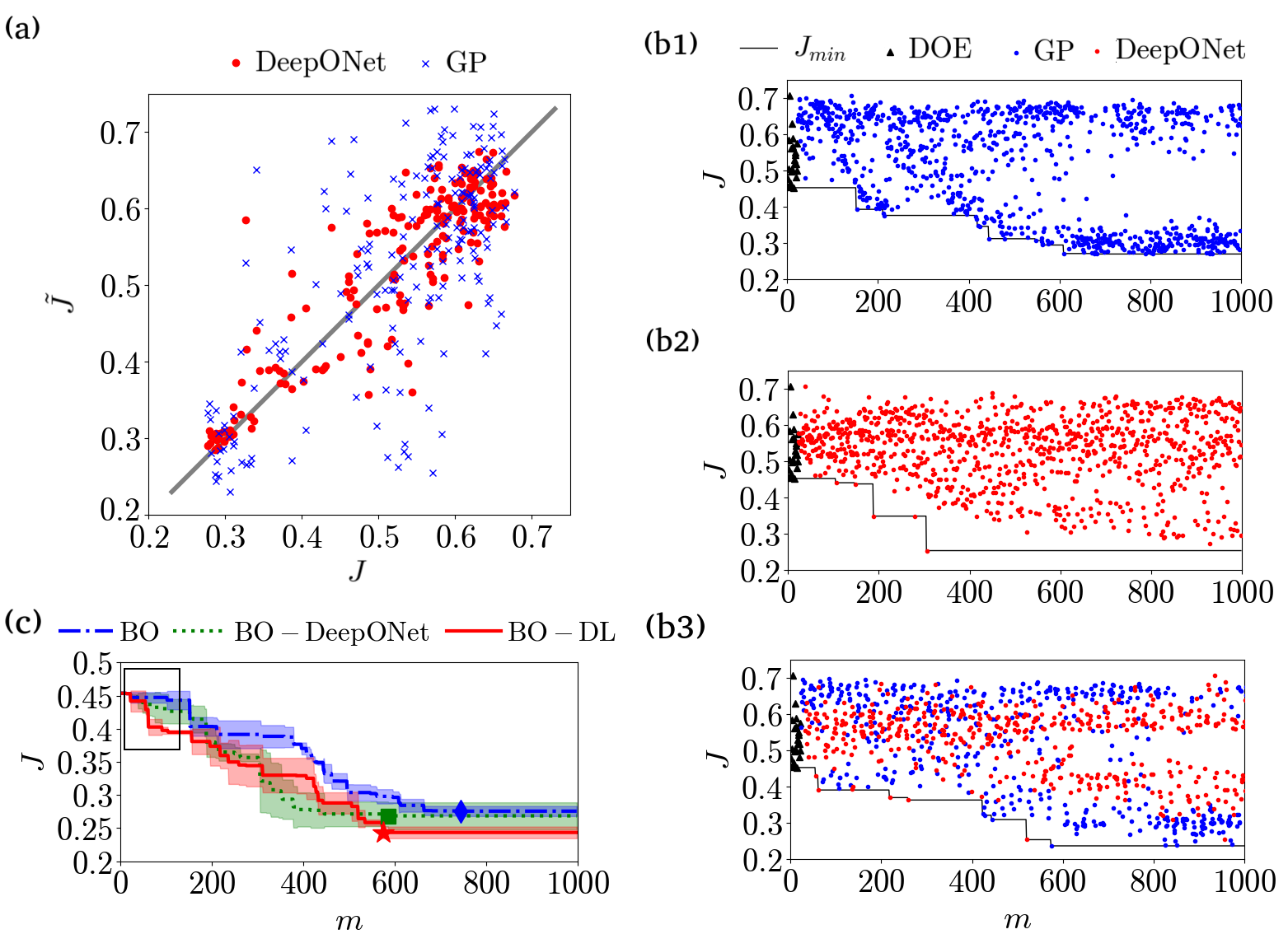}
        \caption{(a) Prediction of $J$ by GP (cross) and DeepONet (dot). 
        Learning curves of $J_{\rm min}$ using BO (b1), BO-DeepONet (b2) and BO-DL (c).
        (d) Average learning curves of Bayesian optimizers with/without deep leaning.}
        \label{fig:opt_bos}
    \end{figure}
    We first study the capability of the surrogate model, GP, and DeepONet, to predict the cost function $J$ as a response to the excitation input $\mathbf{b}$.
    Then, the Bayesian optimizers based on each of the two surrogate models (BO and BO-DeepONet) are tested on our plant.
    Finally, the performance of the proposed method deep learning-enhanced Bayesian optimization (BO-DL) in \ref{ToC:method:bo} is illustrated.

    A $k$-fold cross-validation training ($k=5$) of GP and DeepONet model is performed over $1000$ data points with $80/20$ train/test split.
    The data are extracted randomly out of the database from realizations of Bayesian optimizers.
    Figure~\ref{fig:opt_bos}(a) shows the prediction $\tilde{J}(\mathbf{b})$ versus the truth $J(\mathbf{b})$ obtained. 
    The distribution of points along the diagonal shows that DeepONet achieves a lower prediction error than GP.
    This is further explained by the correlation coefficient $R=0.89$ for DeepONet, and $0.71$ for GP.
    The average error of the $k$ tests is measured by the mean squared error (MSE).
    The MSE for GP model is $0.01$, $1\%$ of the range of $J$ value.
    The prediction of DeepONet model is superior, with a MSE equal to $0.005$.
    The learning process of Bayesian optimizer with GP (BO) and DeepONet (BO-DeepONet) is shown in figure~\ref{fig:opt_bos}(b1-2). 
    In figure~\ref{fig:opt_bos}(b1), the learning curve of BO displays a plateau after the initial samples (triangles).
    After $160$ samples, new optima are found and followed by continuous exploitation of the samples near the learning curve.
    The final solution is reached with $J=0.274$ after $745$ evaluations.
    When DeepONet is employed (figure~\ref{fig:opt_bos}b2),
    a better solution  $J=0.256$ is found quickly within $300$ samples.
    This may be attributed to DeepONet's capibility to generalize better for previously unseen data than GP as the cross-validation indicates  \citep{lu2021nmi}.
    After $m=300$, the newly tested parameters cover the entire range of $J$, but no further improvement is observed in the learning curve.
    This suggests that the optimizer focuses on exploration of the search space rather than exploitation like BO.
    Based on the above observations, a joint surrogate model is proposed for this study to combine the advantages of GP in local exploitation and DeepONet in exploring new minima. 
    The Bayesian optimizer based on this new model is described in \S~\ref{ToC:method:bo} and referred to as BO-DL.
    The learning process of BO-DL is given in figure~\ref{fig:opt_bos}(b3) with the samples queried by GP and DeepONet.
    As indicated by the data points on the learning curve, the queries made by DeepONet (red dots) discovers a new minimum with significant reduction of $J$, and GP (blue dots) continues to descend.
    The best solution is obtained at $J=0.237$ within $600$ evaluations.

    The average performance of three Bayesian optimizers above is further studied. Each optimizer is employed for three realizations with a fixed budget of $1000$ evaluations.
    Figure~\ref{fig:opt_bos}(c) reports the average value of the current optimum $J_{min}$ from each optimizer with the standard deviation (shaded region) of three runs. 
    The learning curve starts from $J_{min}=0.45$, the lowest cost value after initialization of $23$ samples, including the unforced case and the other $22$ controlled cases from Latin hypercube sampling in the search spaces.
    The unforced case ($J=1$) is omitted for better visibility of the data.
    The maximum cost of the controlled flow is around $0.7$.
    With around $750$, $580$ and $570$ queries, the average lowest costs $J_{min}$ achieved by BO, BO-DeepONet, and BO-DL are $J=0.274$ (diamond), $J=0.268$ (square), and $J=0.237$ (star), respectively.
    On average,
    BO-DeepONet shows the fastest learning speed (dotted line) but with the largest variation.
    This is owing to DeepONet's capability of predicting the potential minima with a small generalization error.
    Although the descent of BO is the slowest, the optimal results of the three realizations are consistent.
    This indicates that GP provides better interpolation around the minima than DeepONet due to its deterministic nature.
    By combining GP and DeepONet, BO-DL not only demonstrates a comparable learning speed as BO-DeepONet but also inherits the small variance of the final solution from BO. Finally, among the three optimizers, BO-DL derives the best solution.
    In addition,
    the warm-up phase during queries $0$ to $100$ appears to be significantly shortened, denoted by the rectangle in figure~\ref{fig:opt_bos}(c).

    The computational cost of the BO loop is also noteworthy.
    With BO, 
    the computation of the posterior costs $\mathcal{O}(N^3)$,  where $N$ is the number of observations \citep{williams2006mitbook}.
    This makes the algorithm quite slow, even after only a few hundred observations.
    The experience of this study shows that the computation of BO 
    increases from $10$ CPU-seconds to $600$ CPU-seconds after $1000$ iterations on an AMD EPYC 7742 (2.25GHz) processor.
    The BO-DeepONet procedure scales much more favorably. Initially, the first iteration takes $120$ CPU-seconds, increasing only to $180$ CPU-seconds after $1000$ iterations.
    The combination of the two models in BO-DL compromises the cost to an average level.

    For the high-dimensional physical problem,
    we show that Bayesian optimization can benefit not only from more accurate surrogate models
    but also from combining the advantages of parametric and non-parametric predictors.
    The proposed BO-DL holds a fast convergence and efficient exploration with a GP-DeepONet surrogate model.
    Compared with GP, the proposed surrogate model can provide more accurate predictions by leveraging the hidden functional input with DeepONet and scales better as both data size and dimensionality increase.
    In addition, a comparison between the Bayesian methods and bio-inspired approaches is given in Appendix 
    \ref{ToC:appendix}. 
    It is shown that the optimizer with a surrogate model employed, particularly a deep-learning model, shows more advantage in the current problem.
    \subsection{Exploration and characterization of the search space}
    \label{ToC:res:maps}
    \begin{figure}[htb]
        \centering
        \includegraphics[width=0.8\textwidth]{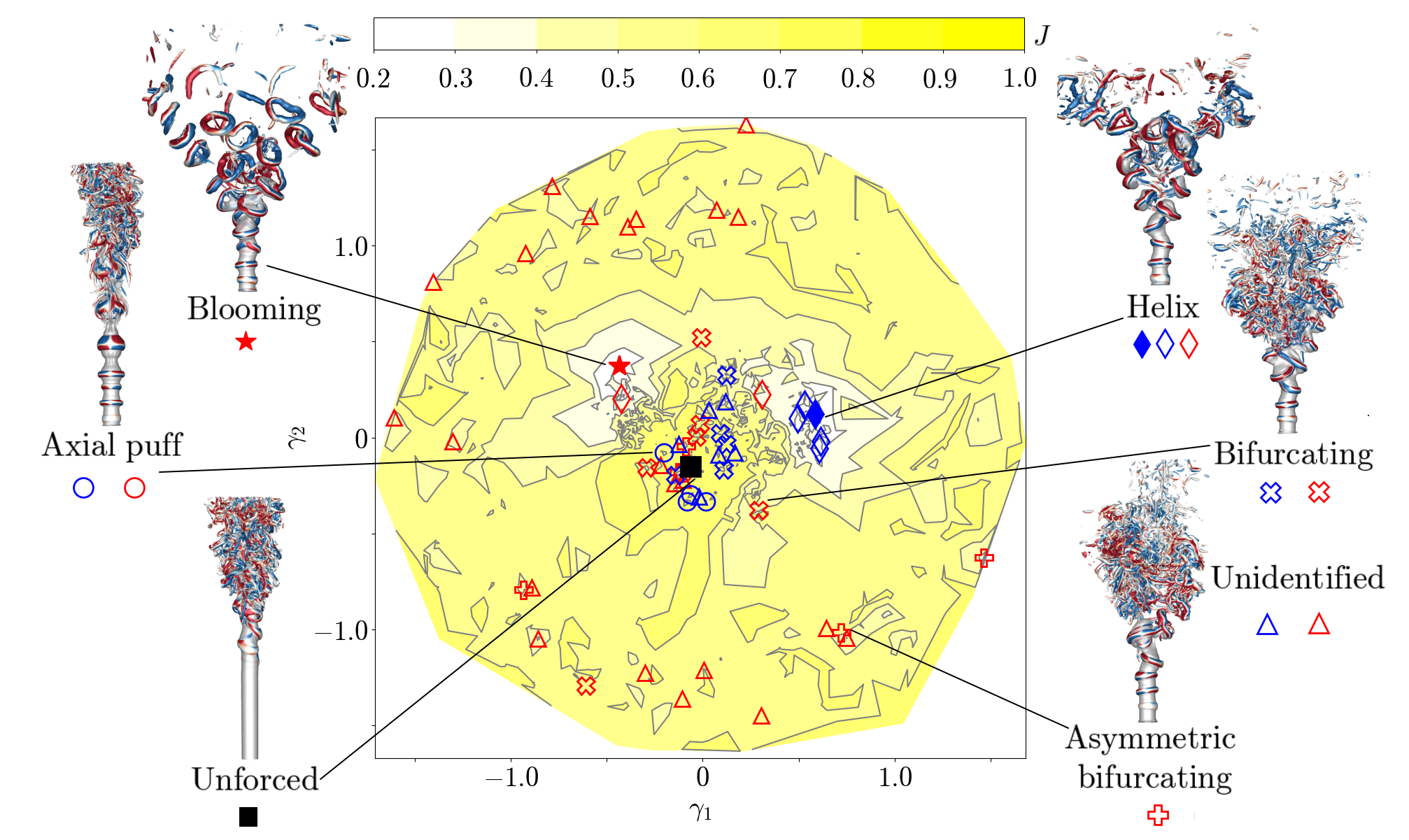}
        \caption{Learning process of BO and BO-DL on the proximity map.
        The unforced case (filled square), the local minima (unfilled symbols), and the final solutions explored by BO (blue-filled diamond) and BO-DL (red-filled star) are highlighted, with related jet patterns.
        }
        \label{fig:opt_map}
    \end{figure}
    In this section, we explore the learning processes of the BO and BO-DL in the $22$-dimensional space with  persistent data topology  \citep{WangTY2023book,WangTY2023pf}.
    This data analysis identifies the cost function minima and their depth,
    i.e.\ their persistence to  noise, and was  inspired by \cite{Edelsbrunner2008proc}.
    The analysis includes  
    the identification of local minima in the high-dimensional actuation space,
    a dimension reduction to a two-dimensional proximity map and corresponding data visualization, and as shown in figure \ref{fig:opt_map}.
    The $22$-dimensional data obtained by both BO and BO-DL are projected on a 2D proximity map  by classical multidimensional scaling. 
    The feature coordinates $\gamma_{ij}$ are chosen to optimally preserve the dissimilarity between control parameters defined by the Euclidean distance $D_{ij}=|\mathbf{b}_i-\mathbf{b}_j|$.
    The map features two large basins of attraction with low values of $J$, as well as small basins distributed around the border.
    A point $\mathbf{b}_0$ is supposed as a local minimum $\mathbf{m}$,
    if there exists a neighborhood $\mathcal{B}$ of  $\mathbf{b}_0$ that satisfies $J(\mathbf{b}_0) \leq \min_{\mathbf{b}\in\mathcal{B}} J(\mathbf{b})$.
    $\mathcal{B}$ is an open set which should include $K$ nearest neighbours of   $\mathbf{b}_0$ measured by Euclidean distance, $K\geq N_D+1$.
    Note that the local minima are assumed based on the obtained discrete data and may change with additional data.
    A total of $57$ local minima are extracted from the data, with $36$ found by BO-DL and $21$ by BO.
    In the proximity map, the unforced case is represented by a black square where both algorithms begin.
    The other symbols denote the derived minima $\mathbf{m}$ found by BO (blue) and BO-DL (red).
    The final BO and BO-DL solutions highlighted by the filled diamond ($J=0.27$) and the filled star ($J=0.24$) are located in the large basins of attractions.
    Most of the minima queried by BO are located in the center of the map, whereas BO-DL also explores outward regions.
    Forced by the control commands corresponding to these minima, different jet patterns are observed, corroborated with the control modes. 
    The axial puff (circles) are close to the unforced case.
    The bifurcating type (cross) distributes widely in the cost range.
    The lower the $J$ value is, the closer to the helix (filled diamond) basin.
    The jets bifurcating to one side are away from the center, surrounded by the other unidentified patterns (triangles).
    Helix (diamond) and blooming (star) jets feature the most substantial performance,
    but the latter is only detected by BO-DL.
    Among the $20$ minima explored by BO, the identified patterns include $6$ helix, $5$ flapping, and $4$ axial puff.
    Among the $37$ minima explored by BO-DL, the identified patterns include $6$ flapping, $5$ asymmetric flapping, $2$ helix, $1$ blooming, and $1$ axial puff.
    In addition, most ($22$) of the $27$ unidentified patterns are detected by BO-DL.

    BO-DL explores not only more minima than BO but also more diverse flow patterns beneficial to the mixing.
    This is probably owing to DeepONet's capability to extrapolate the mapping from the high-dimensional actuation to the mixing response more accurately.
    Two solutions with large basins of attractions in the search space are revealed --- the optimal solution with a $7$-armed blooming jet generated, and the suboptimal with a double-helix shape.
    \subsection{Discussion of the optimized solutions}
    \label{ToC:res:solutions}
    \begin{figure}[htb!]
    \includegraphics[width=\textwidth]{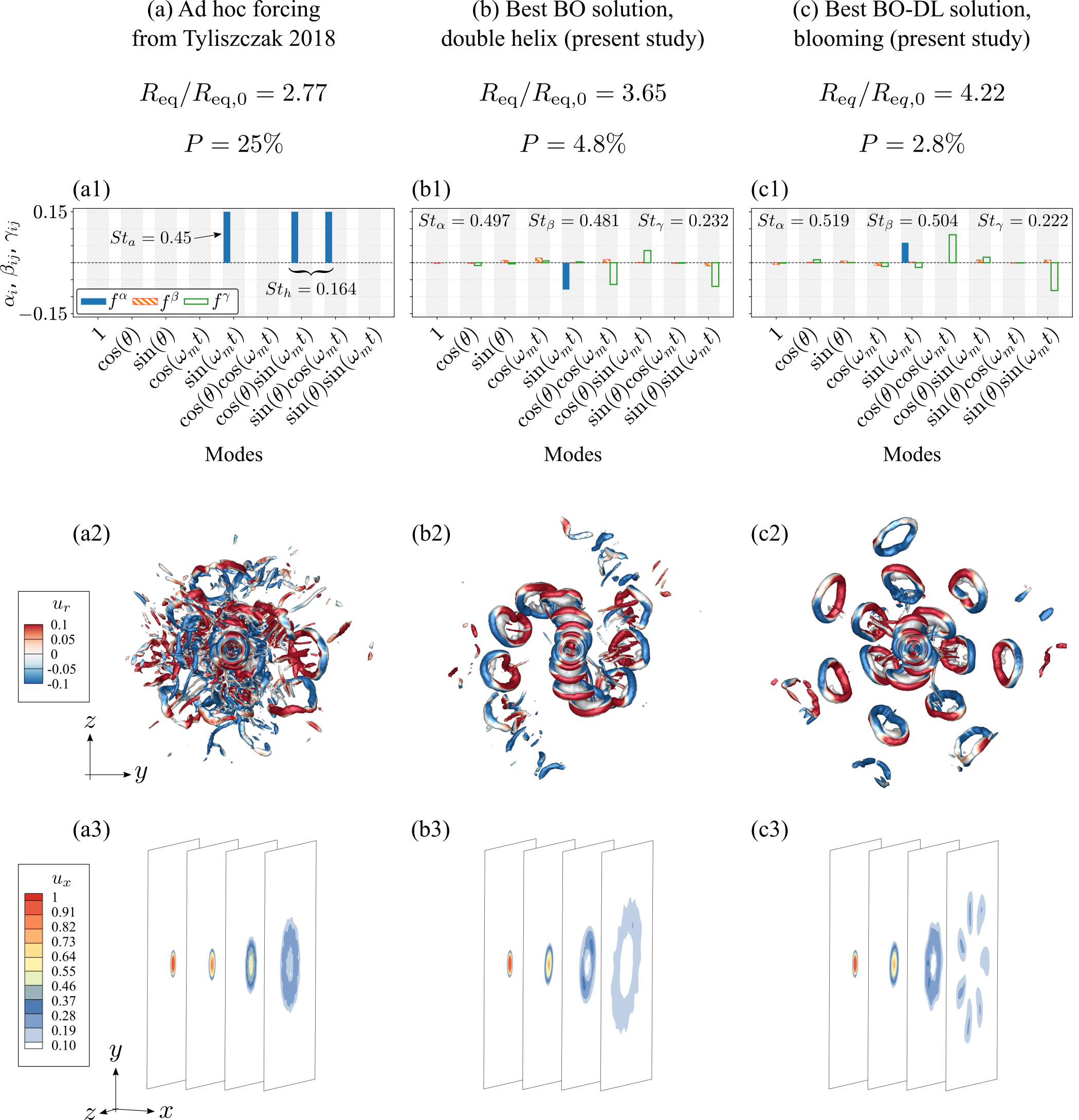}
        \caption{
        Comparison between three mixing jet solutions:
        (a) the ad hoc solution with the best mixing in \cite{Tyliszczak2018ijhff},
        (b) the best solution learned with BO,
        and (c) the best solution learned with BO-DL.
        Top --- forcing modes with the associated mixing and actuation metric. Middle --- (contour online) bottom view of instantaneous isosurfaces of $Q=0.5$ colored by the radial velocity $u_r$. Bottom --- (contour online) contour plots of the streamwise velocity on cross-sectional planes at $x=2D$, $4D$, $6D$, $8D$.
        }
        \label{fig:PerformanceGraphs}
    \end{figure}
    Here, we include three solutions for the discussion: 
    an ad hoc forcing with the best mixing in \citet{Tyliszczak2018ijhff}, BO optimized solution, and the optimal solution of BO-DL.
    The forcing command, the instantaneous snapshots, and the mean flow fields are presented in figure~\ref{fig:PerformanceGraphs}.
    The forcing commands are expressed by the operators in an order of constant, spatial-periodic, temporal-periodic, and traveling waves in figure~\ref{fig:PerformanceGraphs}(a1),~\ref{fig:PerformanceGraphs}(b1) and~\ref{fig:PerformanceGraphs}(c1).
    The axial excitation combining axisymmetric and helical modes has been widely employed to study the bifurcating and blooming jets since \cite{lee1985reporttf}.
    A parametric study of the blooming jets with this type of excitation was performed in \citet{Tyliszczak2018ijhff}, under the same Reynolds number as this study.
    Among various multi-armed jets, the one with 11 arms led to the best mixing performance.
    The excitation was imposed on the axial velocity and combined the axisymmetric mode with Strouhal number {$St_a=0.45$} and the helical mode with {$St_h=0.164$} at the same amplitude, $15\%$ the bulk jet velocity (figure~\ref{fig:PerformanceGraphs}a1).
    The BO solution contains mainly the axisymmetric mode at an amplitude of $8\%$ with $St_\alpha=0.497$ for the bulk, a helical mode at an amplitude of $7\%$, and a flapping mode at an amplitude of $4\%$ with $St_\gamma=0.232$ for radial components in the periphery.
    After removing the expansions with negligible amplitudes, less than $1\%$ of the bulk jet, the control law approximately reads
    \begin{equation}
        \begin{aligned}\label{eq:expr_BO}
    f^{\alpha}(t) \quad  &= 0.08 \sin(2 \pi \times 0.497 U_{\rm j} \>  t/D) \\
    f^{\beta}(\theta,t) &\approx  0 \\
    f^{\gamma}(\theta,t) & \approx  0.04 \cos(\theta) \sin(2 \pi \times 0.232 U_{\rm j} \> t/D) \\
        & \quad -0.07 \cos(\theta- 2 \pi \times 0.232 U_{\rm j} \> t/D).
        \end{aligned}
    \end{equation}
    Because the removed terms hold an amplitude lower than the turbulence intensity at the jet outlet, the approximation hardly changes the flow patterns, with the relative cost difference being less than $1\%$.
    The BO-DL solution contains mainly the axisymmetric mode at an amplitude of $6\%$ with $St_\alpha=0.519$ for the bulk, a helical mode at an amplitude of $8\%$ with $St_\gamma=0.223$ for radial components in the periphery.
    The simplified control law reads
    \begin{equation}\label{eq:expr_BODL}
        \begin{aligned}
        f^{\alpha}(t) \quad &= 0.06 \sin(0.519 t) \\
        f^{\beta}(\theta,t) &\approx  0 \\
        f^{\gamma}(\theta,t) & \approx 0.08 \cos(\theta+2 \pi \times 0.223 U_{\rm j} \> t/D).
        \end{aligned}
    \end{equation}

    Two significant factors to be noted are the axisymmetric forcing Strouhal number $St_\alpha$, and the frequency ratio between the axial and helical modes, {$\alpha=St_\alpha/St_\gamma$.}
    For both BO and BO-DL solutions,
    the axisymmetric forcing Strouhal number falls into the range {$0.4 \lesssim St_\alpha \lesssim 0.6$} to observe bifurcating and blooming jets,
    and coincides with around {$St_\alpha = 0.5$} where the peak spreading occurs \citep{lee1985reporttf, shaabani2020jfm, gohil2015jfm}.
    The BO-DL actuation takes a frequency ratio of $2.34$, which very well agrees with a theoretically derived value $\alpha=7/3$ \cite{tyliszczak2015pof,gohil2015jfm}.
    Interestingly, the ratio of the BO solution which produces a helix jet ($\alpha=2.14$) also falls into this range.
    Moreover, 
    different from the ad hoc excitation using only the axial forcing,
    the radial component in the periphery plays an important role in solutions optimized by both BO and BO-DL.
    \cite{shaabani2020jfm} also concludes radial forcing is the dominant component of helical modes to maximize the spreading angle of a bifurcating jet.
    We extend the importance of radial forcing to the jet spreading globally.
    From an estimate of the momentum flux, 
    the solutions in this study take only $2.8\%$ (BO-DL) and $4.8\%$ (BO) of the main jet, one order lower than the ad hoc excitation ($25\%$).
    One reason is the low amplitudes, and another is the forcing applied into the local boundary region (see \S~\ref{ToC:setup:b}) rather than the whole jet, which leads to a more efficient control.
    This represents the physical reality of small actuators installed on the wall of the inlet nozzle, like flap arrays in \cite{suzuki1999symposium}, only affecting the boundary layers.

\begin{figure}
        \centering
        \includegraphics[width=0.8\textwidth]{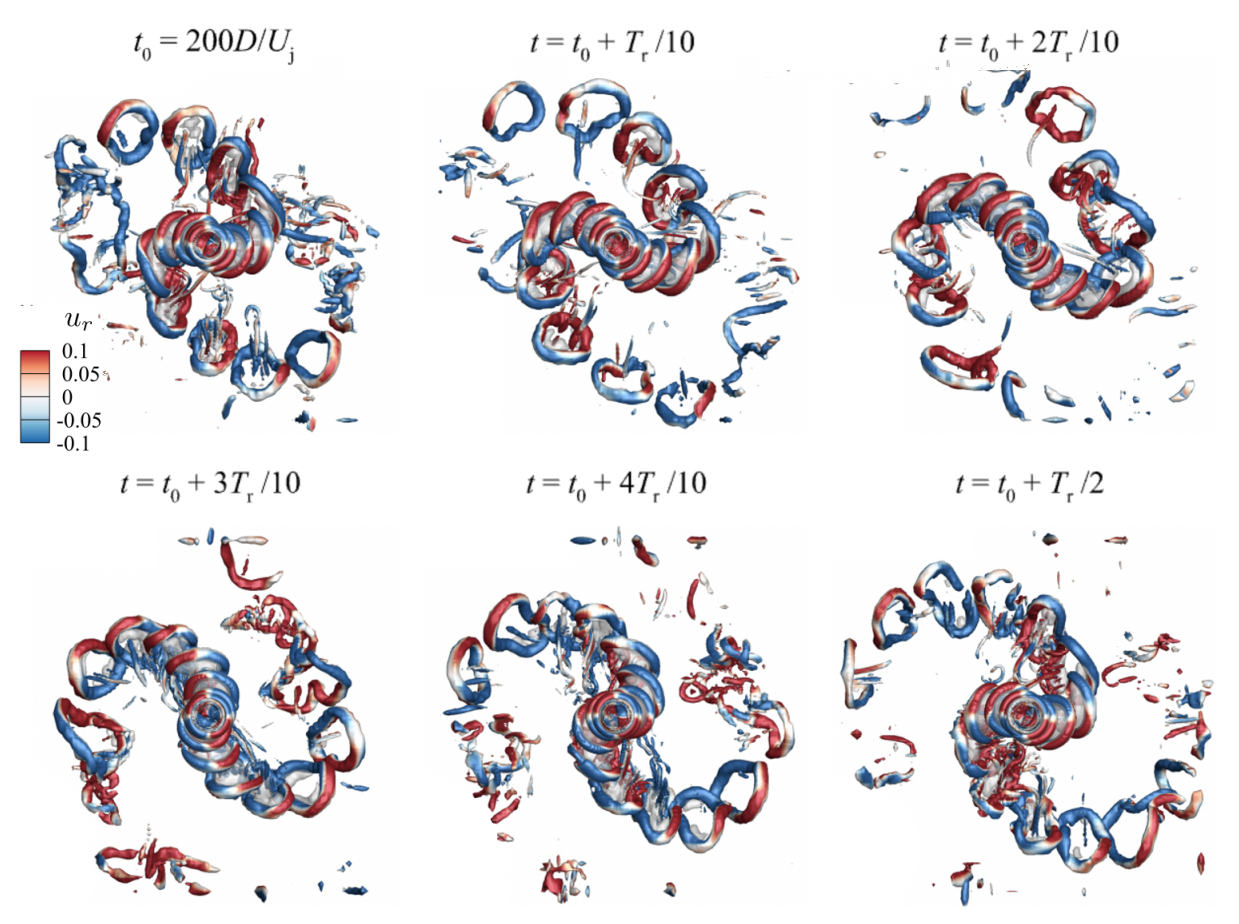}
        \caption{Instantaneous isosurfaces of Q-parameter ($Q=0.5$) for the helical jet, colored by the radial velocity, in a half period of the arm rotation. $T_r = 62.5 D/U_{\rm j}$.}
        \label{fig:res:rotation}
    \end{figure}

    The flow structures are presented by the bottom view of the instantaneous Q-parameter isosurfaces (figure~\ref{fig:PerformanceGraphs}a2, b2, and c2).
    The arms of the ad hoc blooming jet are not explicitly observed due to the interaction between the closely aligned vortex rings.
    A double-helix jet is formulated by the BO solution. The jet bifurcates into two branches, which rotate with a specific frequency (see figure~\ref{fig:res:rotation}) and then experiences continuous bifurcation along the azimuth until the vortex rings break. This type of jet has not been reported in the literature so far.
    We reserve it for future investigation.
    The BO-DL optimized jet produces a $7$-armed blooming jet, 
    with the vortex rings eventually propagating along $7$ different trajectories.
    The contour slices of the time-averaged streamwise velocity also confirm the spreading observed from the vortex rings.
    The 11 branches generated by the ad hoc forcing can be traced in $x=8D$.
    The BO-optimized jet shows more continuous distribution along the circumference due to the azimuthal bifurcation of two helix-shaped arms.
    The blooming jet features the earliest and furthest spreading.
    This leads to the largest effective mixing radius, $4.22 R_{\rm eq,0}$ at $x=8D$,
    followed by BO optimized jet with $R_{\rm eq}=3.65 R_{\rm eq,0}$,
    and the ad hoc forced jet with $R_{\rm eq}=2.77 R_{\rm eq,0}$.

    \begin{figure}
        \centering
        \includegraphics[width=0.9\textwidth]{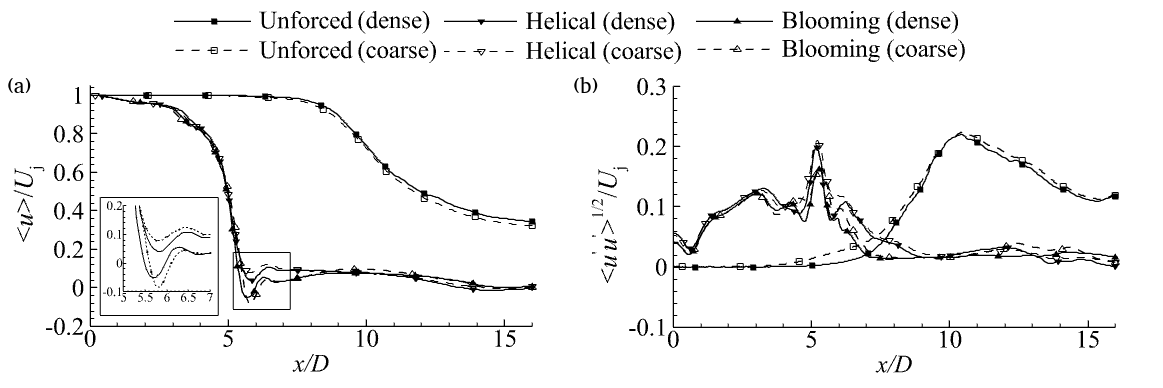}
        \caption{Mean profiles (a) and fluctuations (b) of the axial velocity for the unforced flow, the helical (BO solution), and the blooming jet (BO-DL solution). The solid and dashed lines denote the results calculated separately by the coarse and dense meshes.}
        \label{fig:res:ucl}
    \end{figure}
    \begin{figure}
        \centering
        \includegraphics[width=\textwidth]{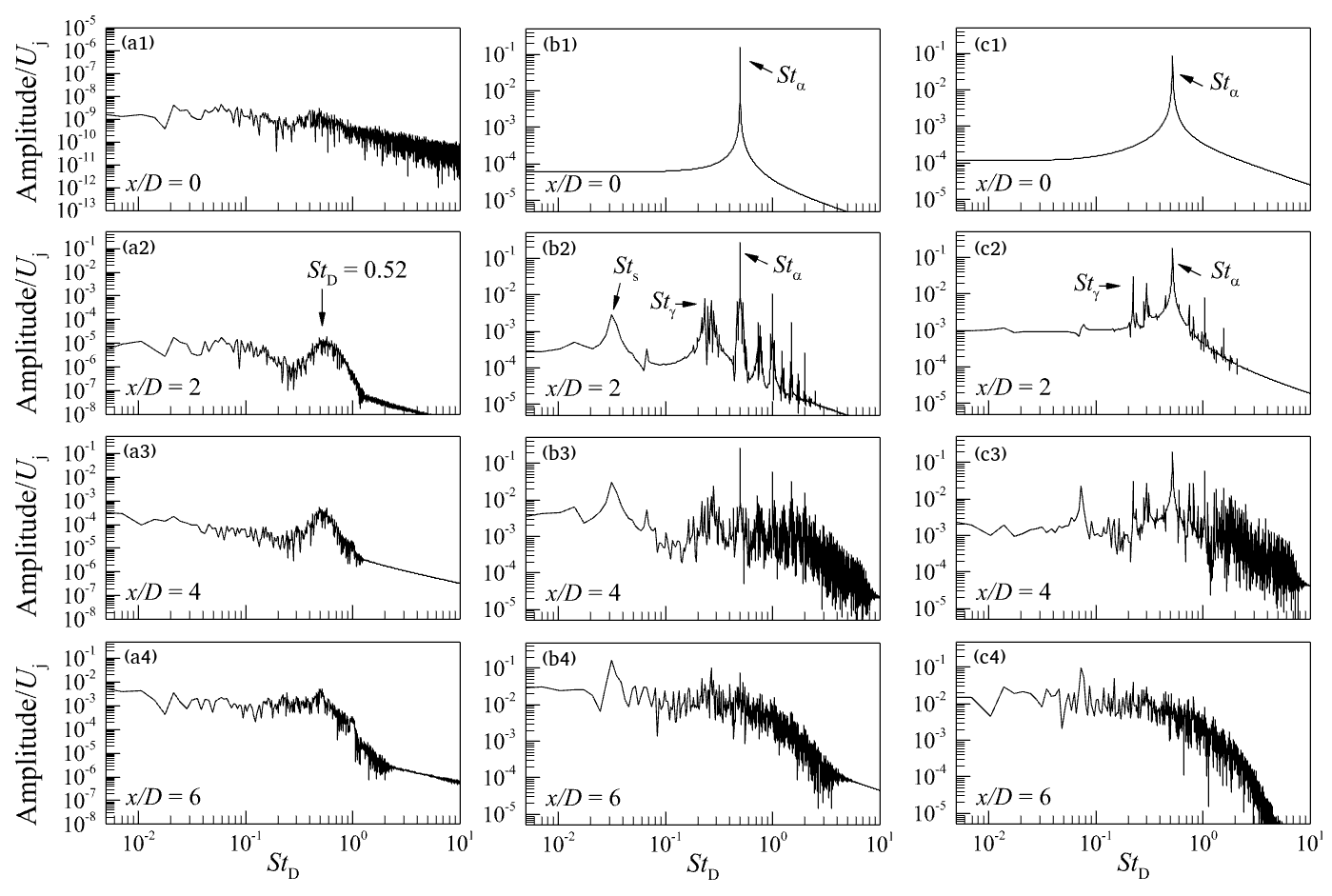}
        \caption{Axial velocity spectrum for the unforced (a1-a3), the helical (b1-b3), and the blooming jet (c1-c3) at $x=0D$, $2D$, $4D$, and $6D$.}
        \label{fig:res:st}
    \end{figure}
    Figure~\ref{fig:res:ucl} shows the axial profiles of the time-averaged centerline velocity and its fluctuation for the unforced and forced jets. 
    The results obtained on the coarse and dense meshes agree well, except for slight discrepancies in the region $5D< x< 8D$.
    Compared with the unforced case in figure~\ref{fig:res:ucl}(a), the length of the potential core shortens significantly from $7.5D$ to $2D$ for both helical and blooming jets.
    Beyond the potential core, 
    the velocity drops steeply and even reaches small negative values in the blooming jet. 
    As a similar behavior observed by
    \cite{Tyliszczak2018ijhff}, this is the effect of jet splitting resulting in a local pressure drop. 
    As a result, the reversal flow amplifies the local turbulence intensity 
    to the peak at $x\approx 5D$, as indicated in figure~\ref{fig:res:ucl}(b).
    The initial fluctuation level in both jets corresponds to the imposed forcing amplitudes of the bulk forcing term. A small decrease at $x<1D$ is caused by the lack of energy in a range of low-wave numbers \citep{Kempf_et_al_FTaC_2005}. 
    For the controlled jets, the waves of the fluctuation profiles around the peak are attributed to the interactions between the Kelvin-Helmholtz instability and the forcing disturbance.
    Further downstream, the fluctuations drop nearly to zero along with a low mean velocity. 
    The fluctuation profile for the unforced jet shows a very low turbulence level until $x\approx 7D$,
    and then slowly increases to the maximum around $x=10D$. 

    Figure \ref{fig:res:st} shows 
    the amplitude spectra of the centerline velocity at four localizations along the axis, $x=0D$, $2D$, $4D$, and $6D$.
    These results are presented versus the Strouhal number $St_D= \omega D/2 \pi U_{\rm j}$. The spectrum of the unforced jet is nearly flat at the inlet as the imposed turbulent signal does not contain any characteristic frequency. 
    The high-frequency components ($St_D>1$) are dampened downstream, and a broadband peak emerges around $St_D=0.52$. This falls within the range of the preferred mode frequency $St_p=0.3-0.64$ \citep{crow1971jfm,GutmarkHo_PoF_1983,Sadeghi_Pollard_PoF_2012}. 
    Note that the optimal $St_\alpha$ predicted by BO-DL for the blooming jet perfectly matches the current preferred mode $St_p = 0.52$. 
    This finding is consistent with previous studies \citep{Tyliszczak2014ftc,gohil2015jfm,Tyliszczak2018ijhff} which concluded that the jet splitting phenomenon is most pronounced when $St_\alpha$ is equal to $St_p$.
    The initial spectra of the helix and blooming jets characterize a distinct peak at $St_\alpha$.
    The peaks related to the helical forcing $St_\gamma$ can be observed from $x=2D$.
    The high-frequency harmonics also appear due to the interactions between generated toroidal vortices. 
    In the case of the helical jet, the peak at $St_D\approx0.032$ is also noteworthy. 
    We find that this frequency coincides with the azimuthal motion of the helical arms, with a period  $T_r$ equal to $62.5 D/U_{\rm j}$.
    Figure~\ref{fig:res:rotation} shows the snapshots of the helical jet, depicting the positions of its arms during the period of $31.25 D/U_{\rm j}$, which corresponds to the detected $St_D\approx0.032$.
    The relationship between the frequency of the rotation and the one associated with forcing terms is left for future study.

\section{Conclusions and outlook}
    \label{ToC:conclusion}
    We perform a global optimization of the jet control modes, parameterized in a $22$-dimensional search space. 
    The forcing includes axial and radial components that are defined to approximate a general periodic function of time and azimuthal angle.
    The design space allows the actuation to emulate various forcing modes that have been studied.
    This high-dimensional problem for jet mixing improvement is tackled by Bayesian Optimization (BO).
    We advance BO by incorporating a deep-learning enhanced surrogate model.
    This surrogate model combines the non-parametric method GP for fast local descent and the parametric method Deep Operator Network (DeepONet) for efficient exploration of the search space.
    The proposed optimizer BO-DL is more efficient in searching for minima and more scalable to large datasets.
    To further understand the optimized high-dimensional solutions,
    we propose a topological analysis of the optimization data.  
    The achieved control landscape features two persistent (pronounced) minima,
    a global minimum corresponding to a 7-armed blooming jet generated, and a suboptimal parameter with a double-helix shape that performs comparably.
    Intriguingly, many of the less persistent minima also correspond to known actuated jet mixing mechanisms.

    Compared with the unforced jet, both the helical and the blooming jet shorten the length of the potential core substantially from $7.5D$ to $2D$. 
    The valley of the mean centerline velocity is located around $6D$ in the downwash, corresponding to the peak of the fluctuation profiles.
    The reversal flow in the blooming jet amplifies the local turbulence intensity,
    and leads to even negative velocity in the centerline.
    Both of the optimized control laws show the radial component dominates the non-axisymmetric forcing mode.
    The optimized forcing for helical jet is a triple-mode that combines the axisymmetric bulk component, a helical, and a flapping mode in the periphery.
    The $7$-armed blooming jet is produced by a dual-mode forcing with only axisymmetric and helical modes.
    The better performance of the latter is attributed to the exact match between the axisymmetric forcing Strouhal number and the preferred mode frequency found by BO-DL.
    The forced flows are characterized by a distinct peak at the Strouhal number of the axisymmetric mode, and the effect of the helical forcing appears later.
    Intriguingly,
    a peak at the low Strouhal number in the helical jet coincides with the azimuthal motion of the helical arms.

    This study emphasizes the importance of effective exploration for machine learning-based optimization in flow control, particularly in high-dimensional design spaces.
    The proposed BO-DL enhances the explorative feature of BO by improving the model accuracy and increasing the solvable model capacity.
    Therefore, BO-DL can serve as an alternative to classical BO when there is a need for greater complexity.
    In addition to parallelizing GP and DeepONet in the Bayesian framework, we can also incrementally increase the model complexity.
    For example,
    we can use the controller obtained by GP to accelerate the learning process of the DeepONet.
    Furthermore,
    DeepONet can also be employed as a function approximator like DRL, which deserves future study under a different framework --- Bayesian experimental design.
    As an add-on, 
    the proposed persistent data topology analysis can help to characterize the control landscape  from the discrete data produced by different optimizers.
    Persistent data minima indicate literature known and unknown mixing mechanisms.
    Finally,
    we expect the proposed BO-DL and topological data analysis for effective learning and characterization of the search spaces could contribute to more flow control problems.
\section*{Acknowledgements}
    This work is supported by the National Natural Science Foundation of China under grants 12172109 and 12302293, by the Guangdong Basic and Applied Basic Research Foundation under grant 2022A1515011492,  and by the Shenzhen Science and Technology Program under grant JCYJ20220531095605012. This work is also supported by the Polish National Science Center under grant 2018/31/B/ST8/00762.
    We gratefully acknowledge inspiring and formative contributions of Antoine Blanchard and Themis Sapsis to the employed Bayesian optimization algorithm and its interpretation.
    We express our sincere thanks to Zhutao Jiang
    for testing our optimization algorithmss in a companion jet mixing experiment and deepening our physical insights and 
    to Thomas Weise for enriching our roadmap on optimization algorithms
    and to Herbert Edelsbrunner and Hans-Christian Hege for a discussion of topological data analyses.
    \section*{Declaration of interests}
    The authors report no conflict of interest.
\begin{appendix}
\section{Comparison of Bayesian optimization with two bio-inspired optimizers}
\label{ToC:appendix}
    \begin{figure}[h]
        \centering
        \includegraphics[width=0.5\textwidth]{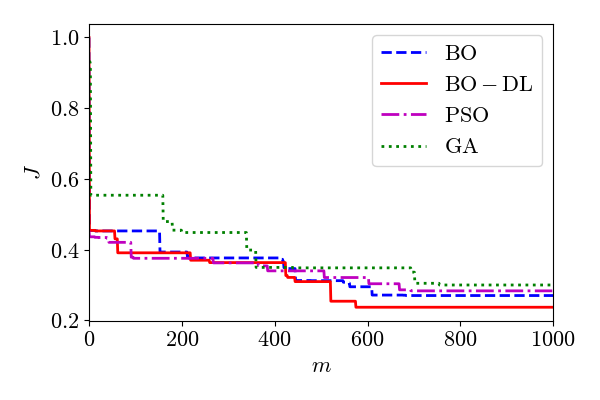}
        \caption{Learning curves of $J_{\rm min}$ using BO, BO-DL, PSO and GA.}
        \label{fig:curve_bo_pso_ga}
    \end{figure}
    \begin{figure}[h]
        \includegraphics[width=\textwidth]{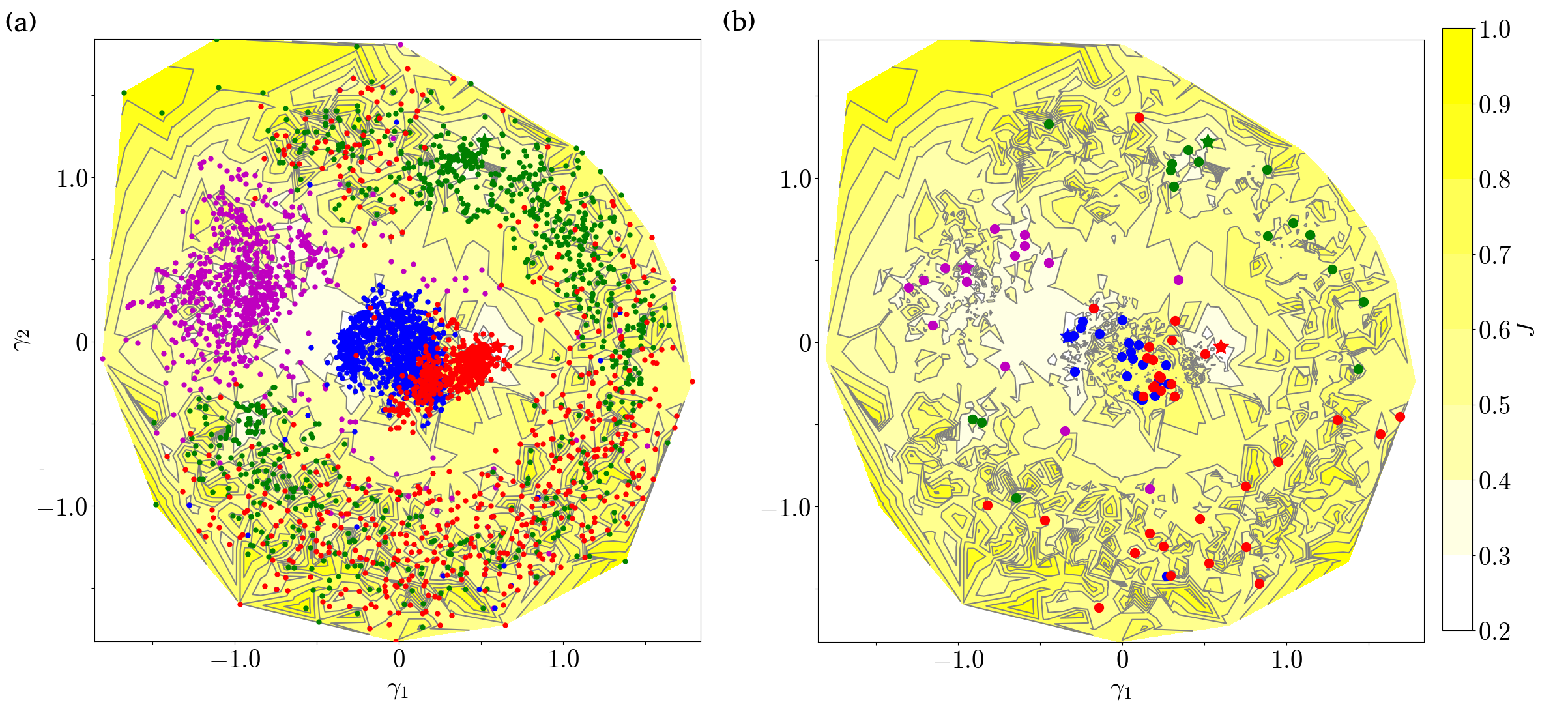}
        \caption{Landscape with visited data (a) and derived local minima (b) of BO (blue), BO-DL (red), PSO (magenta) and GA (green) from unknown in the whole space.}
        \label{fig:map_bo_pso_ga}
    \end{figure}
    The Bayesian optimizers, BO and BO-DL, employed in this study, have been compared with two popular biologically inspired methods \citep{wahde2008book}:  Particle Swarm Optimization (PSO) and Genetic Algorithm (GA).
    Here, a variant of PSO, Particle Swarm Optimization through Targeted, 
    Position‑Mutated, Elitism (PSO-TPME) is employed \citep{Shaqarin2023ijcis},
    and GA is realized following \cite{wright1991proceeding}.
    Figure \ref{fig:curve_bo_pso_ga} 
    displays the learning curve of one realization of the four methods. The learning curves give an indication of the learning speed of each method.
    PSO converges to a solution with the cost $J=0.283$ slightly higher than BO ($J=0.273$),
    and GA ends with a even higher cost $J=0.3$.
    Figure \ref{fig:map_bo_pso_ga} presents all the evaluated points (a), and the local minima (b) derived from the combined database.
    The tested solutions during the search are denoted by the colored dots in figure \ref{fig:map_bo_pso_ga} (a).
    The derived minima are denoted by the filled circles with corresponding colors in figure \ref{fig:map_bo_pso_ga} (b). 
    The converged solutions are depicted by stars.
    PSO and GA fall into the local minima in the upper left and the upper right corner, respectively.
    Interestingly,
    the search process of these methods show different features.
    PSO moves all the particles (magenta dots) towards the best region detected. 
    Finally, all particles accumulate in the upper left region and get stuck.
    The minima (magenta circles) 
    are found along the direction of gradient descent.
    GA searches the minima in one neighborhood but is extremely inefficient in exploring further regions.
    Most of the exploration away from the global minimum in the right upper region in figure \ref{fig:curve_bo_pso_ga} (a) ends with no local minima in figure \ref{fig:curve_bo_pso_ga} (b).
    Owing to the prediction by GP, BO converges to the region with a lower cost quickly.
    Moreover,
    with the deep-learning enhanced surrogate model,
    BO-DL not only obtains the best minimum (red star) but also reveals more potential minima in a wider neighbourhood (red circles).
    In high-dimensional search spaces, exploration based on accurate estimators is more efficient than random exploration.
    For the current problem, 
    the optimizer based  on a surrogate model, particularly a deep-learning model, shows more advantages than bio-inspired optimizers.
\end{appendix}
\bibliographystyle{jfm}
\bibliography{Main}

\begin{thebibliography}{51}
\expandafter\ifx\csname natexlab\endcsname\relax\def\natexlab#1{#1}\fi
\def\au#1{#1} \def\ed#1{#1} \def\yr#1{#1}\def\at#1{#1}\def\jt#1{\textit{#1}}
  \def\bt#1{#1}\def\bvol#1{\textbf{#1}} \def\vol#1{#1} \def\pg#1{#1}
  \def\publ#1{#1}\def\arxiv#1{#1}\def\org#1{#1}\def\st#1{\textit{#1}}

\bibitem[Ball {\em et~al.\/}(2012)Ball, Fellouah \& Pollard]{ball2012pas}
{\sc \au{Ball, C.~G.}, \au{Fellouah, H.} \& \au{Pollard, A.}} \yr{2012}
  \at{The flow field in turbulent round free jets}.  \jt{Prog. Aerosp. Sci.}
  \bvol{50},  \pg{1--26}.

\bibitem[Blanchard \& Sapsis(2021)]{blanchard2021jcp}
{\sc \au{Blanchard, A.} \& \au{Sapsis, T.}} \yr{2021}  \at{Bayesian
  optimization with output-weighted optimal sampling}.  \jt{J. Comp. Phys.}
  \bvol{425},  \pg{109901}.

\bibitem[Blanchard {\em et~al.\/}(2021)Blanchard, Cornejo~Maceda, Fan, Li,
  Zhou, Noack \& Sapsis]{blanchard2021ams}
{\sc \au{Blanchard, A.~B.}, \au{Cornejo~Maceda, G.~Y.}, \au{Fan, D.}, \au{Li,
  Y.}, \au{Zhou, Y.}, \au{Noack, B.~R.} \& \au{Sapsis, T.~P.}} \yr{2021}
  \at{Bayesian optimization for active flow control}.  \jt{Acta Mech. Sin.}
  \pg{pp. 1--3}.

\bibitem[Boguslawski {\em et~al.\/}(2019)Boguslawski, Wawrzak \&
  Tyliszczak]{boguslawski2019jfm}
{\sc \au{Boguslawski, A.}, \au{Wawrzak, K.} \& \au{Tyliszczak, A.}} \yr{2019}
  \at{A new insight into understanding the crow and champagne preferred mode: a
  numerical study}.  \jt{J.\ Fluid Mech.}  \bvol{869},  \pg{385--416}.

\bibitem[Brunton {\em et~al.\/}(2020)Brunton, Noack \&
  Koumoutsakos]{Brunton2020arfm}
{\sc \au{Brunton, S.~L.}, \au{Noack, B.~R.} \& \au{Koumoutsakos, P.}} \yr{2020}
   \at{Machine learning for fluid mechanics}.  \jt{Ann.~Rev.~Fluid Mech.}
  \bvol{52},  \pg{477--508}.

\bibitem[Corke \& Kusek(1993)]{corke1993jfm}
{\sc \au{Corke, T.~C.} \& \au{Kusek, S.~M.}} \yr{1993}  \at{Resonance in
  axisymmetric jets with controlled helical-mode input}.  \jt{J. Fluid Mech.}
  \bvol{249},  \pg{307--336}.

\bibitem[{Cornejo Maceda} {\em et~al.\/}(2021){Cornejo Maceda}, Li, Lusseyran,
  Morzy\'nski \& Noack]{Cornejo2021jfm}
{\sc \au{{Cornejo Maceda}, G.~Y.}, \au{Li, Y.}, \au{Lusseyran, F.},
  \au{Morzy\'nski, M.} \& \au{Noack, B.~R.}} \yr{2021}  \at{Stabilization of
  the fluidic pinball with gradient-based machine learning control}.  \jt{J.
  Fluid Mech.}  \bvol{917},  \pg{A42:1--43}.

\bibitem[Crow \& Champagne(1971)]{crow1971jfm}
{\sc \au{Crow, S.~C.} \& \au{Champagne, F.~H.}} \yr{1971}  \at{Orderly
  structure in jet turbulence}.  \jt{J.\ Fluid Mech.}  \bvol{48}~(3),
  \pg{547--591}.

\bibitem[Danaila \& Boersma(2000)]{danaila2000pf}
{\sc \au{Danaila, I.} \& \au{Boersma, B.~J.}} \yr{2000}  \at{Direct numerical
  simulation of bifurcating jets}.  \jt{Phys. Fluids}  \bvol{12}~(5),
  \pg{1255--1257}.

\bibitem[Duriez {\em et~al.\/}(2017)Duriez, Brunton \& Noack]{Duriez2017book}
{\sc \au{Duriez, T.}, \au{Brunton, S.~L.} \& \au{Noack, B.~R.}} \yr{2017} {\em
  Machine Learning Control --- Taming Nonlinear Dynamics and Turbulence\/},
  \st{Fluid Mechanics and Its Applications},  \vol{vol. 116}.
  \publ{Springer-Verlag}.

\bibitem[Edelsbrunner \& Harer(2008)]{Edelsbrunner2008proc}
{\sc \au{Edelsbrunner, H.} \& \au{Harer, J.}} \yr{2008}  \at{Persistent
  homology --- a survey}.  \bt{In {\em Surveys on Discrete and Computational
  Geometry: Twenty Years Later\/} (ed. \ed{J.~E. Goodman, J.~Pach \&
  R.~Pollack})}, ,  \vol{vol. 458},  \pg{pp. 257--282}.  \publ{AMS Bookstore}.

\bibitem[Gohil {\em et~al.\/}(2015)Gohil, Saha \& Muralidhar]{gohil2015jfm}
{\sc \au{Gohil, T.~B.}, \au{Saha, A.~K.} \& \au{Muralidhar, K.}} \yr{2015}
  \at{Simulation of the blooming phenomenon in forced circular jets}.  \jt{J.
  Fluid Mech.}  \bvol{783},  \pg{567--604}.

\bibitem[Guastoni {\em et~al.\/}(2023)Guastoni, Rabault, Schlatter \&
  Azizpour]{Guastoni2023epje}
{\sc \au{Guastoni, L.}, \au{Rabault, J.}, \au{Schlatter, P.} \& \au{Azizpour,
  H.~Vinuesa, R.}} \yr{2023}  \at{Deep reinforcement learning for turbulent
  drag reduction in channel flows}.  \jt{Eur. Phys. J. E}  \bvol{46},  \pg{27}.

\bibitem[Gutmark \& Ho(1983)]{GutmarkHo_PoF_1983}
{\sc \au{Gutmark, E.} \& \au{Ho, C.M.}} \yr{1983}  \at{{Preferred modes and the
  spreading rates of jets}}.  \jt{{Phys. Fluids}}  \bvol{26},  \pg{2932--2938}.

\bibitem[Hilgers \& Boersma(2001)]{hilgers2001fdr}
{\sc \au{Hilgers, A.} \& \au{Boersma, B.~J.}} \yr{2001}  \at{Optimization of
  turbulent jet mixing}.  \jt{Fluid Dyn. Res.}  \bvol{29}~(6),  \pg{345}.

\bibitem[Hussain \& Zaman(1980)]{hussain1980jfm}
{\sc \au{Hussain, A. K. M.~F.} \& \au{Zaman, K. B. M.~Q.}} \yr{1980}
  \at{Vortex pairing in a circular jet under controlled excitation. {P}art 2.
  {C}oherent structure dynamics}.  \jt{J. Fluid Mech.}  \bvol{101}~(3),
  \pg{493--544}.

\bibitem[Jordan \& Colonius(2013)]{Jordan2013arfm}
{\sc \au{Jordan, P.} \& \au{Colonius, T.}} \yr{2013}  \at{Wave packets and
  turbulent jet noise}.  \jt{Annu. Rev. Fluid Mech.}  \bvol{45}~(1),
  \pg{173--195}.

\bibitem[Kempf {\em et~al.\/}(2005)Kempf, Klein \&
  Janicka]{Kempf_et_al_FTaC_2005}
{\sc \au{Kempf, A.}, \au{Klein, M.} \& \au{Janicka, J.}} \yr{2005}
  \at{Efficient generation of initial- and inﬂow-conditions for transient
  turbulent flows in arbitrary geometries}.  \jt{Flow Turbul. Combust.}
  \bvol{74},  \pg{67--84}.

\bibitem[Koumoutsakos {\em et~al.\/}(2001)Koumoutsakos, Freund \&
  Parekh]{Koumoutsakos2001aiaaj}
{\sc \au{Koumoutsakos, P.}, \au{Freund, J.} \& \au{Parekh, D.}} \yr{2001}
  \at{Evolution strategies for automatic optimization of jet mixing}.  \jt{AIAA
  J.}  \bvol{39}~(5),  \pg{967--969}.

\bibitem[Lee \& Reynolds(1985)]{lee1985reporttf}
{\sc \au{Lee, M.} \& \au{Reynolds, W.~C.}} \yr{1985}  \bt{Bifurcating and
  blooming jets.} {\em Tech. Rep.\/}.  \org{Thermosciences Division, Department
  of Mechanical Engineering, Stanford University}.

\bibitem[Lu {\em et~al.\/}(2021)Lu, Jin, Pang, Zhang \& Karniadakis]{lu2021nmi}
{\sc \au{Lu, L.}, \au{Jin, P.}, \au{Pang, G.}, \au{Zhang, Z.} \&
  \au{Karniadakis, G.~E.}} \yr{2021}  \at{Learning nonlinear operators via
  {D}eep{ON}et based on the universal approximation theorem of operators}.
  \jt{Nat. Mach. Intell.}  \bvol{3}~(3),  \pg{218--229}.

\bibitem[Mankbadi \& Liu(1981)]{mankbadi1981ptrs}
{\sc \au{Mankbadi, R.} \& \au{Liu, J. T.~C.}} \yr{1981}  \at{A study of the
  interactions between large-scale coherent structures and fine-grained
  turbulence in a round jet}.  \jt{Philos. Trans. Royal Soc. A}
  \bvol{298}~(1443),  \pg{541--602}.

\bibitem[Nair \& Goza(2023)]{nair2023jfmr}
{\sc \au{Nair, N.~J.} \& \au{Goza, A.}} \yr{2023}  \at{Bio-inspired
  variable-stiffness flaps for hybrid flow control, tuned via reinforcement
  learning}.  \jt{J. Fluid Mech.}  \bvol{956},  \pg{R4}.

\bibitem[Nathan {\em et~al.\/}(2006)Nathan, Mi, Alwahabi, Newbold \&
  Nobes]{nathan2006prog}
{\sc \au{Nathan, G.~J.}, \au{Mi, J.}, \au{Alwahabi, Z.~T.}, \au{Newbold, G.
  J.~R.} \& \au{Nobes, D.~S.}} \yr{2006}  \at{Impacts of a jet's exit flow
  pattern on mixing and combustion performance}.  \jt{Prog. Energy Combust.
  Sci.}  \bvol{32}~(5),  \pg{496--538}.

\bibitem[Parekh(1989)]{parekh1989phd}
{\sc \au{Parekh, D.~E.}} \yr{1989} {\em Bifurcating jets at high Reynolds
  numbers\/}.  \publ{Stanford University}.

\bibitem[Pickering {\em et~al.\/}(2022)Pickering, Guth, Karniadakis \&
  Sapsis]{pickering2022ncs}
{\sc \au{Pickering, E.}, \au{Guth, S.}, \au{Karniadakis, G.~E.} \& \au{Sapsis,
  T.~P.}} \yr{2022}  \at{Discovering and forecasting extreme events via active
  learning in neural operators}.  \jt{Nat. Comput. Sci.}  \bvol{2}~(12),
  \pg{823--833}.

\bibitem[Pino {\em et~al.\/}(2023)Pino, Schena, Rabault \& Mendez]{pino2023jfm}
{\sc \au{Pino, F.}, \au{Schena, L.}, \au{Rabault, J.} \& \au{Mendez, M.~A.}}
  \yr{2023}  \at{Comparative analysis of machine learning methods for active
  flow control}.  \jt{J. Fluid Mech.}  \bvol{958},  \pg{A39}.

\bibitem[Rabault {\em et~al.\/}(2019)Rabault, Kuchta, Jensen, R{\'e}glade \&
  Cerardi]{rabault2019jfm}
{\sc \au{Rabault, J.}, \au{Kuchta, M.}, \au{Jensen, A.}, \au{R{\'e}glade, U.}
  \& \au{Cerardi, N.}} \yr{2019}  \at{Artificial neural networks trained
  through deep reinforcement learning discover control strategies for active
  flow control}.  \jt{J.\ Fluid Mech.}  \bvol{865},  \pg{281--302}.

\bibitem[Sadeghi \& Pollard(2012)]{Sadeghi_Pollard_PoF_2012}
{\sc \au{Sadeghi, H.} \& \au{Pollard, A.}} \yr{2012}  \at{{Effects of passive
  control rings positioned in the shear layer and potential core of a turbulent
  round jet}}.  \jt{Phys. Fluids}  \bvol{24},  \pg{115103}.

\bibitem[Shaabani-Ardali {\em et~al.\/}(2020)Shaabani-Ardali, Sipp \&
  Lesshafft]{shaabani2020jfm}
{\sc \au{Shaabani-Ardali, L.}, \au{Sipp, D.} \& \au{Lesshafft, L.}} \yr{2020}
  \at{Optimal triggering of jet bifurcation: an example of optimal forcing
  applied to a time-periodic base flow}.  \jt{J. Fluid Mech.}  \bvol{885},
  \pg{A34}.

\bibitem[Shahriari {\em et~al.\/}(2015)Shahriari, Swersky, Wang, Adams \&
  De~Freitas]{shahriari2015ieee}
{\sc \au{Shahriari, B.}, \au{Swersky, K.}, \au{Wang, Z.}, \au{Adams, R.~P.} \&
  \au{De~Freitas, N.}} \yr{2015}  \at{Taking the human out of the loop: A
  review of {B}ayesian optimization}.  \jt{Proc. IEEE}  \bvol{104}~(1),
  \pg{148--175}.

\bibitem[Shaqarin \& Noack(2023)]{Shaqarin2023ijcis}
{\sc \au{Shaqarin, T.} \& \au{Noack, B.~R.}} \yr{2023}  \at{A fast-converging
  particle swarm optimization through targeted, position-mutated, elitism
  ({PSO}-{TPME})}.  \jt{Int. J. Comput. Intell. Syst.}  \bvol{16},  \pg{6}.

\bibitem[da~Silva \& M{\'e}tais(2002)]{da2002pf}
{\sc \au{da~Silva, C.~B.} \& \au{M{\'e}tais, O.}} \yr{2002}  \at{Vortex control
  of bifurcating jets: A numerical study}.  \jt{Phys. Fluids}  \bvol{14}~(11),
  \pg{3798--3819}.

\bibitem[Sonoda {\em et~al.\/}(2023)Sonoda, Liu, Itoh \&
  Hasegawa]{Sonoda2023jfm}
{\sc \au{Sonoda, T.}, \au{Liu, Z.}, \au{Itoh, T.} \& \au{Hasegawa, Y.}}
  \yr{2023}  \at{Reinforcement learning of control strategies for reducing skin
  friction drag in a fully developed turbulent channel flow}.  \jt{J. Fluid
  Mech.}  \bvol{960},  \pg{A30}.

\bibitem[Suzuki {\em et~al.\/}(1999)Suzuki, Kasagi \&
  Suzuki]{suzuki1999symposium}
{\sc \au{Suzuki, H.}, \au{Kasagi, N.} \& \au{Suzuki, Y.}} \yr{1999} Active
  control of an axisymmetric jet with an intelligent nozzle.  \bt{In {\em First
  Symposium on Turbulence and Shear Flow Phenomena\/}}. Begel House Inc.

\bibitem[Tyliszczak(2014)]{tyliszczak2014jcp}
{\sc \au{Tyliszczak, A.}} \yr{2014}  \at{A high-order compact difference
  algorithm for half-staggered grids for laminar and turbulent incompressible
  flows}.  \jt{J.\ Comp.\ Phys.}  \bvol{276},  \pg{438--467}.

\bibitem[Tyliszczak(2015{\natexlab{{\em a\/}}})]{tyliszczak2015cf}
{\sc \au{Tyliszczak, A.}} \yr{2015{\natexlab{{\em a\/}}}}  \at{{LES}-{CMC}
  study of an excited hydrogen flame}.  \jt{Combust. Flame}  \bvol{162}~(10),
  \pg{3864--3883}.

\bibitem[Tyliszczak(2015{\natexlab{{\em b\/}}})]{tyliszczak2015pof}
{\sc \au{Tyliszczak, A.}} \yr{2015{\natexlab{{\em b\/}}}}  \at{Multi-armed
  jets: A subset of the blooming jets}.  \jt{Phys.\ Fluids}  \bvol{27}~(4),
  \pg{041703}.

\bibitem[Tyliszczak(2018)]{Tyliszczak2018ijhff}
{\sc \au{Tyliszczak, A.}} \yr{2018}  \at{Parametric study of multi-armed jets}.
   \jt{Int. J. Heat Fluid Flow}  \bvol{73},  \pg{82--100}.

\bibitem[Tyliszczak \& Geurts(2014)]{Tyliszczak2014ftc}
{\sc \au{Tyliszczak, A.} \& \au{Geurts, B.~J.}} \yr{2014}  \at{Parametric
  analysis of excited round jets-numerical study}.  \jt{Flow Turbul. Combust.}
  \bvol{93},  \pg{221--247}.

\bibitem[Utkin {\em et~al.\/}(2006)Utkin, Keshav, Kim, Kastner, Adamovich \&
  Samimy]{utkin2006jfd}
{\sc \au{Utkin, Y.~G.}, \au{Keshav, S.}, \au{Kim, J.~H.}, \au{Kastner, J.},
  \au{Adamovich, I.~V.} \& \au{Samimy, M.}} \yr{2006}  \at{Development and use
  of localized arc filament plasma actuators for high-speed flow control}.
  \jt{J. Phys. D: Appl. Phys.}  \bvol{40}~(3),  \pg{685}.

\bibitem[Vignon {\em et~al.\/}(2023{\natexlab{{\em a\/}}})Vignon, Rabault,
  Vasanth, Alcántara-Ávila, Mortensen \& Vinuesa]{Vignon2023pof_marl}
{\sc \au{Vignon, C.}, \au{Rabault, J.}, \au{Vasanth, J.},
  \au{Alcántara-Ávila, F.}, \au{Mortensen, M.} \& \au{Vinuesa, R.}}
  \yr{2023{\natexlab{{\em a\/}}}}  \at{{Effective control of two-dimensional
  Rayleigh–Bénard convection: Invariant multi-agent reinforcement learning
  is all you need}}.  \jt{Phys. Fluids}  \bvol{35}~(6),  \pg{065146}.

\bibitem[Vignon {\em et~al.\/}(2023{\natexlab{{\em b\/}}})Vignon, Rabault \&
  Vinuesa]{Vignon2023pof}
{\sc \au{Vignon, C.}, \au{Rabault, J.} \& \au{Vinuesa, R.}}
  \yr{2023{\natexlab{{\em b\/}}}}  \at{{Recent advances in applying deep
  reinforcement learning for flow control: Perspectives and future
  directions}}.  \jt{Phys. Fluids}  \bvol{35}~(3),  \pg{031301}.

\bibitem[Wahde(2008)]{wahde2008book}
{\sc \au{Wahde, M.}} \yr{2008} {\em Biologically inspired optimization methods:
  an introduction\/}.  \publ{WIT press}.

\bibitem[Wang {\em et~al.\/}(2023{\natexlab{{\em a\/}}})Wang, {Cornejo~Maceda}
  \& Noack]{WangTY2023book}
{\sc \au{Wang, T.}, \au{{Cornejo~Maceda}, G.~Y.} \& \au{Noack, B.~R.}}
  \yr{2023{\natexlab{{\em a\/}}}} {\em xPDT: A Toolkit for Persistent Data
  Topology\/}.  \publ{Universit\"atsbibliothek der Technischen Universit\"at
  Braunschweig, Germany.}

\bibitem[Wang {\em et~al.\/}(2023{\natexlab{{\em b\/}}})Wang, Yang, Chen, Li,
  Iollo, {Cornejo~Maceda} \& Noack]{WangTY2023pf}
{\sc \au{Wang, T.}, \au{Yang, Y.}, \au{Chen, X.}, \au{Li, P.}, \au{Iollo, A.},
  \au{{Cornejo~Maceda}, G.~Y.} \& \au{Noack, B.~R.}} \yr{2023{\natexlab{{\em
  b\/}}}}  \at{Topologically assisted optimization for rotor design}.
  \jt{Phys. Fluids}  \bvol{35},  \pg{055105}.

\bibitem[Wawrzak {\em et~al.\/}(2015)Wawrzak, Boguslawski \&
  Tyliszczak]{Wawrzak_Boguslawski_Tyliszczak_FTaC_2015}
{\sc \au{Wawrzak, K.}, \au{Boguslawski, A.} \& \au{Tyliszczak, A.}} \yr{2015}
  \at{{LES} predictions of self-sustained oscillations in homogeneous density
  round free jet}.  \jt{Flow Turbul. Combust.}  \bvol{95},  \pg{437--459}.

\bibitem[Williams \& Rasmussen(2006)]{williams2006mitbook}
{\sc \au{Williams, C.~K.} \& \au{Rasmussen, C.~E.}} \yr{2006} {\em Gaussian
  processes for machine learning\/}.  \publ{MIT press}.

\bibitem[Wright(1991)]{wright1991proceeding}
{\sc \au{Wright, A.~H.}} \yr{1991}  \at{Genetic algorithms for real parameter
  optimization}.  \bt{In {\em Foundations of genetic algorithms\/}}, ,
  \vol{vol.~1},  \pg{pp. 205--218}.  \publ{Elsevier}.

\bibitem[Xu \& Zhang(2023)]{Xu2023jfm}
{\sc \au{Xu, D.} \& \au{Zhang, M.}} \yr{2023}  \at{Reinforcement-learning-based
  control of convectively unstable flows}.  \jt{J. Fluid Mech.}  \bvol{954},
  \pg{A37}.

\bibitem[Zhou {\em et~al.\/}(2020)Zhou, Fan, Zhang, Li \& Noack]{zhou2020jfm}
{\sc \au{Zhou, Y.}, \au{Fan, D.}, \au{Zhang, B.}, \au{Li, R.} \& \au{Noack,
  B.~R.}} \yr{2020}  \at{Artificial intelligence control of a turbulent jet}.
  \jt{J. Fluid Mech.}  \bvol{897},  \pg{A27}.

\end{thebibliography}

\end{document}